\documentclass[12pt]{article}
\usepackage{amsmath}
\usepackage{amssymb}
\usepackage{comment}
\usepackage{enumerate}
\usepackage{cite}
\numberwithin{equation}{section}
\title{Love numbers for the Myers--Perry geometry}
\author{Rhucha Deshpande\footnote{rdeshpande@albany.edu}\  \   and Oleg Lunin\footnote{olunin@albany.edu}\\
\ \\
 {\it\normalsize
Department of Physics,}\\ {\it\normalsize University at Albany (SUNY),}\\ 
{\it\normalsize Albany, NY 12222, USA}\\
}
\date{}
\begin{document}
\def\be{\begin{equation}}
\def\bea{\begin{eqnarray}}
\def\ee{\end{equation}}
\def\eea{\end{eqnarray}}
\def\d{\partial}
\def\eps{\varepsilon}
\def\la{\lambda}
\def\La{\Lambda}
\def\b{\bigskip}
\def\nn{\nonumber \\}
\def\p{\partial}
\def\t{\tilde}
\def\h{{1\over 2}}
\def\r{{\rho}}
\def\qb{\bar q}
\def\be{\begin{equation}}
\def\bea{\begin{eqnarray}}
\def\ee{\end{equation}}
\def\eea{\end{eqnarray}}
\def\OO{\mathcal{O}}
\def\aa{\mathfrak{a}}
\def\bb{\mathfrak{b}}
\def\cc{\mathfrak{c}}
\def\b{\bigskip}
\def\u{\uparrow}

\maketitle

\begin{abstract}
\noindent
We analyze Love numbers for scalar fields in the Myers-Perry geometry in arbitrary dimensions. The tidal responses of black holes are encoded in two constants which encode a potential logarithmic running and the Love number. We argue that the parameter specifying the running is not sensitive to the structure of the horizon, and we explicitly compute this parameter for Myers-Perry black holes in arbitrary dimensions by utilizing the asymptotic expansion of the wave equation. The Love numbers rely on the details of the boundary conditions at the horizon, and we compute them using several distinct approximations.
\end{abstract}

\newpage

\tableofcontents

\newpage

\section{Introduction}

Studies of black holes and their excitations have played an important role in advancing our understanding of classical and quantum gravity. Recent breakthroughs in direct observations of astrrophysical black holes via detection of electromagnetic \cite{BHemObs} and gravitational \cite{BHgravObs} waves produced by them relied on a detailed understanding of classical fields in the vicinity of these objects \cite{NumGR1w1,NumGR1w2,NumGR1w3,NumGR1w4,NumGR2w1,NumGR2w2,NumGR2w3,
NumGR2w4,NumGR2w5}. Dynamical excitations are also crucial for gaining insights into quantum properties of black holes, such as Hawking radiation \cite{Hawking}. Furthermore, studies of light excitations of higher--dimensional black holes and branes \cite{PreAdSCFTw1,PreAdSCFTw2,PreAdSCFTw3,PreAdSCFTw4,PreAdSCFTw5,PreAdSCFTw6,
PreAdSCFTw7} have led to significant progress in understanding of these objects and stimulated the discovery of AdS/CFT correspondence \cite{AdSCFTw1,AdSCFTw2,AdSCFTw3}. In recent years, significant efforts have been dedicated to gaining insights into the internal structure of black holes through 
studying tidal deformations of their horizons. In this article we will analyze the tidal deformations of rotating black holes in arbitrary dimensions. 

\bigskip

When a compact object, such as a neutron star or a black hole, is placed in an external gravitational field, the distortion of the object's equilibrium shape causes adjustments in the distribution of mass, which in turn generate induced multipole moments in the gravitational potential. These moments are characterized by the tidal Love numbers (TLN) \cite{LoveOrig}. Historically introduced in the context of Newtonian geophysics to
describe the elasticity of Earth, Love numbers have become crucial parameters in relativistic astrophysics. For example, for neutron stars, the values of the TLN depend on the 
nuclear equation of state, and thus it provides an observational window into the
behavior of matter at very high densities \cite{LoveNP1,LoveSchw2,LoveSchw3,LoveSchw4,LoveNP2,LoveLandryW1,LoveLandryW2}. Tidal responses of black holes have been studied as well \cite{StaticLN1,LoveSchw1,LoveSchw2,LoveSchw3,LoveSchw4}, and in recent years there has been a renewed interest in evaluating Love numbers for these objects and in analyzing the corresponding deformations of the horizon, which provide insights into internal structure of black holes, and which can be detected through direct observations.

In four dimensions, the Love numbers have been computed for scalar excitations of both static \cite{StaticLN1} and rotating \cite{LoveKerrM0,LoveKerrM1,LoveKerrM2,LoveKerrM3} black holes, and these results were later extended to bosonic fields with higher spins  \cite{LoveKerrM3,StatLNgraviton}, and to fermions \cite{LoveKerrM4f}. Interestingly, all Love numbers turned out to be zero, and this protection relied on the hidden $SL(2,R) \times U(1)$ symmetry of all dynamical equations, which was analyzed in \cite{LoveSym0,LoveKerrM3}. Evaluation of the Love numbers and investigation of the symmetry were based on a remarkable property of both Schwarzschild and Kerr geometries: all dynamical equations admit separation of variables, which allows one to reduce the problem of calculating the tidal responses to the study of ordinary differential equations for functions of radial variable. While this is not surprising for the Schwarzschild case, which has spherical symmetry, separation of variables for Kerr solution relies on another hidden symmetry of the background, which is encoded in irreducible Killing and Killing--Yano tensors \cite{Carter1w1,Carter1w2,Carter1w3}. This symmetry has been used to separate equations for scalars \cite{Carter1w1,Carter1w2}, as well as spinors, vectors, and gravitons \cite{TeukW1,TeukW2}, and it was the resulting radial ODEs that were analyzed in  \cite{LoveKerrM0,LoveKerrM1,LoveKerrM2,LoveKerrM3,LoveKerrM4f}.

\bigskip

Black holes in higher dimensions have also been studied extensively \cite{Tang,MyersPerry,HorStromW1,HorStromW2,HorStromW3,SenW1,
SenW2,SenW3,SenW4,SenW5,SenW6,BMPVw1,BMPVw2,GLPP}, and given their importance for understanding the Bekenstein--Hawking entropy \cite{StrVafa,BMPVw1,BMPVw2} and for exploring strongly coupled field theories \cite{5dRotAdScftW1,5dRotAdScftW2,5dRotAdScftW3,5dRotAdScftW4,5dRotAdScftW5} in the framework of the AdS/CFT correspondence \cite{AdSCFTw1,AdSCFTw2,AdSCFTw3}, it is natural to analyze Love numbers for such geometries. In the static Schwarzschild--Tangherlini case \cite{Tang}, such analysis was performed in \cite{StaticLN2}, where it was shown that the Love numbers could be nontrivial if angular momentum of the excitation took certain values\footnote{Specifically, the Love numbers vanished if and only if the ratio $({\ell}/(d-3))$ was an integer, where $\ell$ is an integer quantum number parameterizing the angular momentum of the excitation, which can be a scalar, vector or tensor, and $d$ is the dimension of spacetime. We will review this construction in section \ref{SecStatic}.}. The symmetries responsible for these results were discussed in \cite{LoveSym0,LoveSym1,LoveSym2,LoveSym3,StatLN3sym}. 
For rotating black holes, such as Myers--Perry (MP) \cite{MyersPerry} or 
Gibbons--\hspace{0pt}Lu--\hspace{0pt}Page--\hspace{0pt}Pope (GLPP) \cite{GLPP} geometries, one encounters two potential problems: 
separability of the dynamical equations is not obvious, and radial dependence of the metric is more complicated than in the four--dimensional case. Remarkably, the first obstacle is avoided due to a large hidden symmetry of the MP and GLPP black holes, which leads to full separation of equations for  scalars \cite{Kub1w1,Kub1w2,Kub1w3,Kub1w4,Kub1w5,Kub1w6,Kub2w1,Kub2w2,Kub2w3,Kub3w1,
Kub3w2,Kub3w3,UniqueKYw1,UniqueKYw2,
UniqueKYw3,UniqueKYw4}\footnote{See also earlier related work  \cite{PreKubW1,PreKubW2,PreKubW3}, as well as extensions of separabilty to charged black holes and black branes \cite{StrYanoW1,LarsenW1,LarsenW2,LarsenW3,LarsenW4,LarsenW5,StrYanoW2,ChLKill}.}, vectors \cite{LMaxw,KubMaxwW1,KubMaxwW2,Proca,KubMaxw1w1,KubMaxw1w2,KubMaxw1w3}, and higher antisymmetric forms \cite{LBfield}\footnote{Equations for gravitons are also separable in several special cases \cite{KodamaRotOneAw1,KodamaRotOneAw2,GWequalA,GWnearHor,AllDimOtherW1,
AllDimOtherW2,AllDimOtherW3,NPallD,LGravWave}.}. Unfortunately, the second obstacle for computing Love numbers, the complicated form of the radial equation, prevents explicit analytical evaluation of tidal responses in dimensions larger than five. In the five dimensional case, the problem reduces to the hypergeometric equation, and relevant Love numbers were evaluated in \cite{RotLN5d,RSST}. Beyond that, Love numbers have been computed only in various approximations for some specific dimensions \cite{LoveRotHd1,LoveRotHd2}. 
Our approach is complementary to this work.

In contrast to articles \cite{LoveRotHd1,LoveRotHd2}, we will keep the dimension of space arbitrary, which will allow us to get insights into the properties of the tidal coefficients as functions of both dimensions and angular quantum numbers of excitations. Specifically, we will clarify the distinction between the Love numbers and the ``logarithmic running'' found in \cite{StaticLN2,RotLN5d,RSST}. While such running does depend on the metric, we will demonstrate that it is fully determined by the first few terms in the multipole expansion at infinity, and thus it does not carry information about the structure of the horizon\footnote{Similar observation in the static case was made in \cite{LoveRunNo1,LoveRunNo2}.}. This structure is encoded in the choice of the integration constant, which is subleading to the running, and we will extend the work of \cite{StaticLN2,RotLN5d,RSST} by computing this constant for the static case and for rotating black holes in five dimensions, when analytical methods are available. Beyond five dimensions, we perform an exact evaluation of the logarithmic running for general MP black holes and use various approximations to extract the genuine Love numbers. Specifically, we will evaluate the tidal coefficients for slowly rotating black holes, and separately we will calculate these numbers in the WKB approximation, assuming that the angular quantum numbers of excitations are large, while the rotation parameters of the black hole remain arbitrary. We will also compute the exact Love numbers for several extremal black holes, which have not been discussed in the past.

\bigskip

This paper has the following organization. 
In section \ref{SecLogRunSec} we formulate the equations studied throughout this article, set up the problems of evaluating the Love numbers and logarithmic runnings for arbitrary black holes, and calculate the running coefficients for the Myers--Perry geometries in all dimensions. In section \ref{SecSlow} we evaluate the Love numbers for slowly rotating black holes using perturbative expansion in rotation parameters. Excitations of  extremal black holes turn out to be rather peculiar, so they are analyzed in section \ref{SecExtremal}, where exact expressions for the Love numbers in five and seven dimensions are obtained. Section \ref{SecWKB} explores the WKB approximation: this time the rotation parameters characterizing black holes are arbitrary, but the angular momenta of excitations are taken to be large. Finally, in section \ref{SecAngles} we analyze perturbations with nontrivial quantum numbers associated with cyclic directions. Some technical details are presented in appendices. 

\section{Love numbers and logarithmic running}
\label{SecLogRunSec}

In this section we will analyze the wave equation in the backgrounds of rotating black holes in arbitrary dimensions. While in the static case, the analytical solutions for radial profiles and explicit expressions for the Love numbers can be found in terms of hypergeometric functions, equations in the rotating case can be solved only in certain dimensions. Nevertheless, one can extract a universal pattern: at large distances from the sources, the coefficient in front of the most interesting power of $r$ may contain a mixture of the constant corresponding to the Love number and a term with logarithmic running. Such running terms have been found for several classes of black holes in the past \cite{StaticLN2,RotLN5d,RSST}, and these constructions used explicit solutions regular at the horizon. By observing that the running terms are completely fixed by the expansion at infinity, we evaluate these contributions for rotating in arbitrary dimensions\footnote{A similar approach to static black holes was advocated in \cite{LoveRunNo1,LoveRunNo2}.}. Calculation of genuine Love numbers, which are highly sensitive to the boundary conditions at the horizon, will be carried out in subsequent sections. 

\bigskip
This section has the following organization. In subsection \ref{SecEqnSetup} we set up the problem which will be studied throughout this article. Specifically, we briefly review the Myers--Perry geometry in arbitrary dimensions and wave equation in this background. In subsection \ref{SecStatic} we review the known facts about Love numbers in static geometries and the origin of the logarithmic running. Subsections \ref{SecLogSpec}, \ref{SecLogSpecEven}, and \ref{SecLogGen} are dedicated to the original analysis of logarithmic running in arbitrary dimensions. Since the structures of the MP black holes differ between odd and even dimensions, we perform detailed investigations of these two cases separately in subsections \ref{SecLogSpec} and \ref{SecLogSpecEven}, before summarizing the general lessons in subsection \ref{SecLogGen}.

\subsection{Scalar fields in the Myers--Perry black geometry}
\label{SecEqnSetup}

Let us consider the wave equation on the background of rotating black holes in arbitrary dimensions. The relevant geometries are reviewed in the Appendix \ref{AppMP}, and the structure of backgrounds differ between the odd-- and even--dimensional cases. Therefore we will analyze these cases separately, and we begin with the odd--dimensional spacetime. 

\bigskip

\noindent
{\bf Odd dimensions}

As reviewed in Appendix \ref{AppMP}, separation of the wave equation in the background of the $d=(2n+1)$--dimensional Myers--Perry black holes with metrics (\ref{MPodd}) leads to the system of ordinary differential equations (\ref{MPodeOdd}). Focusing on vanishing quantum numbers associated with cyclic directions, we conclude the wave equation with separable ansatz
\bea\label{PhiSeparMain}
\nabla^2\Phi=0,\qquad \Psi=\Phi(r)\prod_{i=1}^{n-1}X_i(x_i)
\eea
leads to the system of ODEs
\bea\label{OdeOddMPmain}
r\frac{d}{dr}\left[\frac{\Delta}{r}\frac{d\Phi}{dr}\right]-r^2 P_{n-2}[r^2]\Phi=0,\quad
x_i\frac{d}{dx_i}\left[\frac{H_i}{x_i}\frac{dX_i}{dx_i}\right]-(x_i)^2P_{n-2}[-x_i^2]X_i=0
\eea
Here functions $(\Delta,H_i)$ associated with black hole geometry are given by
\bea
\Delta=\left[\prod_k^n (r^2+a_k^2)\right]-M r^2,\quad 
H_i=\prod_k^n (a_k^2-x_i^2).
\eea
Without loss of generality, we will assume a particular ordering of the rotation parameters, then the ranges of the $x$--coordinates are given by
\bea\label{RangeMPodd}
a_1\le x_1\le a_2\le \dots \le x_{n-1}\le a_n\,.
\eea
To connect equations (\ref{OdeOddMPmain}) with degenerate cases analyzed in other sections, we consider the limit where some rotation parameters become equal:
\bea
a_k=\dots =a_{k+p}\,.
\eea
Given the ranges (\ref{RangeMPodd}), we conclude that some coordinates become degenerate in this limit, so to proceed, we split the rotation parameters using a control parameter $\eps$,
\bea\label{RangeMPexp1}
a_{k+1}=a_k+\eps b_{k+1},\ \dots,\ a_{k+p}=a_k+\eps b_{k+p},
\eea 
re-define the coordinates as
\bea\label{RangeMPexp2}
&&x_k=a_k+\eps y_k,\qquad \dots,\ x_{k+p-1}=a_k+\eps y_{k+p-1},\\
&&0\le y_k\le b_{k+1}\le \dots \le y_{k+p-1}\le b_{k+p},\nonumber
\eea
and send $\eps$ to zero while keeping $(b_s,y_s)$ fixed. Expanding the angular equation with $i=k$ from (\ref{OdeOddMPmain}) in powers of $\eps$, we conclude that there are $(p-1)$ constraints on the polynomial $P_{n-2}$:
\bea\label{PolynReduc}
P_{n-2}[r^2]=(r^2+a_k^2)^{p-1}P_{n-p-1}[r^2].
\eea
Note that this constraint appears only if three or more rotation parameters collide, and we implicitly assumed that $p>0$. If rotation parameters cluster in groups with $(p_1,\dots p_s)$ entries (all $p_i$ are positive), then the polynomial $P_{n-2}$ has the form
\bea
P_{n-2}[r^2]={\tilde P}[r^2]\prod_j (r^2+a_{k_j}^2)^{p_j-1},\quad n\ge \sum (p_j+1).
\eea
In particular, if all rotation parameters are equal, then $p=n-1$ in (\ref{PolynReduc}), and all coefficients in polynomial $P_{n-2}$, apart from the overall normalization, are fixed. This leads to a significant simplification: while generically one has to analyze the boundary problem for the angular equations in (\ref{OdeOddMPmain}) to quantize the coefficients of $P_{n-2}$, in the case of equal rotation all such parameters are fixed by symmetries. This is true even for the overall coefficient. Indeed, writing 
\bea\label{PolynMPaEq}
P_{n-2}[r^2]=\Lambda (r^2+a^2)^{n-2}\,,
\eea
we conclude that the parameter $\Lambda$ is dimensionless. Since mass does not enter the angular equations in (\ref{OdeOddMPmain}), $\Lambda$ must be a function of $a$ only, and thus it has to be $a$--independent on the dimensional grounds. Therefore, using the well-known results for separation constant for spherical harmonics, we find
\bea\label{LamOne}
\Lambda=\ell (\ell+d-3),
\eea
where $\ell$ is an integer. The relevant radial differential equation reads
\bea\label{RadiaMPmainFin}
r\frac{d}{dr}\left[\frac{(r^2+a^2)^n-M r^2}{r}\frac{d\Phi}{dr}\right]-\Lambda r^2(r^2+a^2)^{n-2}\Phi=0\,.
\eea
The Love numbers for this simple case will be analyzed in section \ref{SecLogSpec}, before we turn to the general situation in section \ref{SecLogGen}. We conclude this review of the wave equation in the Myers--Perry geometry by making some comments about the even--dimensional case.

\bigskip

\noindent
{\bf Even dimensions}

The even--dimensional Myers--Perry black holes \cite{MyersPerry,Myers} are reviewed in the Appendix \ref{AppMP}. In $d=2(n+1)$--dimensional space, the metric is given by (\ref{MPeven}), and the separable ansatz (\ref{WaveEqnTwo}) for a massless scalar field leads to the system of ordinary differential equations (\ref{MPodeEven}). As before, we will focus on the case of vanishing quantum numbers associated with cyclic directions, then the ansatz simplifies to (\ref{PhiSeparMain}), and the counterpart of the system (\ref{OdeOddMPmain})  reads
\bea\label{OdeEvenMPmain}
\frac{d}{dr}\left[{\Delta}\frac{d\Phi}{dr}\right]-P_{n-1}[r^2]\Phi=0,\quad
\frac{d}{dx_i}\left[{H_i}\frac{dX_i}{dx_i}\right]+P_{n-1}[-x_i^2]X_i=0
\eea
with functions $(\Delta,H_i)$ are given by
\bea
\Delta=\left[\prod_k^n (r^2+a_k^2)\right]-M r,\quad 
H_i=\prod_k^n (a_k^2-x_i^2)
\eea
The analysis of degenerate cases proceeds as before, and for equal rotation parameters we find
\bea\label{PolynEven}
P_{n-1}[r^2]=(\La_1 r^2+\La_2 a^2)(r^2+a^2)^{n-2}\,.
\eea
The value of $\La_1$ is still given by (\ref{LamOne}) ($\La_1=\ell (\ell+d-3)$), and 
to determine $\La_2$, we need to look at angular equations. Recalling that in the even--dimensional case the coordinates are arranged as 
\bea
0\le x_1\le a_1\le\dots \le x_{n-1}\le a_{n-1}\,,
\eea
we conclude that the easiest way to determine $\la_2$ is to look at equation (\ref{OdeEvenMPmain}) for $x_1$ since the limit of equal rotations is straightforward. Dropping the subscript of $x$, we find
\bea\label{LgndrEqn}
\frac{d}{dx}\left[(a^2-x^2)^n\frac{dX}{dx}\right]-(\La_1 x^2-\La_2 a^2)(a^2-x^2)^{n-2}X=0
\eea
Writing $x=a y$, we arrive at a simpler version of this equation,
\bea
(1-y^2)^2X''-2ny(1-y^2)X'-(\La_1 y^2-\La_2 )X=0,
\eea
and a candidate for the solution regular at $y=\pm 1$ can be expressed in terms of the Legendre function\footnote{We are dropping the second Legendre function $Q$ since it gives singular solutions regardless of the value of parameters.}
\bea\label{LamEven}
&&X=(1-y^2)^{-(n-1)/2}P_{\ell +n-1}^{(q)}(y),\nn
&&\La_1=\ell (\ell+d-3),\quad  \La_2=(\ell+n-1)^2+\ell-q^2
\eea
Noting that equation (\ref{LgndrEqn}) naturally comes from the effective action
\bea
S_X=\int dx (a^2-x^2)^{n-2}\left\{(a^2-x^2)^{2}\left[\frac{dX}{dx}\right]^2+
(\La_1 x^2-\La_2 a^2)X^2\right\}\,,\nonumber
\eea
we conclude that normalizability of the solution implies that parameters $(\ell,q)$ are integers and 
\bea
\ell\ge 0,\quad |q|\le \ell +n-1\,.
\eea
To summarize, we conclude that in the even--dimensional case with equal rotation parameters, the radial equation is 
\bea\label{RadialEqnEvenE}
\frac{d}{dr}\left[\left\{(r^2+a^2)^n-2 M r\right\}\frac{d\Phi}{dr}\right]-(\La_1 r^2+\La_2 a^2)(r^2+a^2)^{n-2}\Phi=0
\eea
with eigenvalues $(\La_1,\La_2)$ given by (\ref{LamEven}). 

\subsection{Love numbers in the static case}
\label{SecStatic}

Let us give a brief review of Love numbers in the static case. The relevant equations have been extensively discussed in the literature \cite{StaticLN1,StaticLN2,LoveSym1,LoveSym2,LoveSym3}, so we will just summarize the main features of the solutions. The radial equation reads
\bea\label{ScalEqnSS}
\frac{d}{dr}\left[\left\{r^{d-2}-M r\right\}\frac{d\Phi}{dr}\right]-\ell (\ell+d-3) r^{d-4}\Phi=0
\eea
Since the coefficient in front of the highest derivative vanishes at the horizon located at 
$r=r_h$ with
\bea
r_h\equiv M^{\frac{1}{d-3}}\,,
\eea
only one solution remains regular at this location, and it reads
\bea\label{PhiHyper}
\Phi=c_1 r^{\ell}F\left[-\frac{\ell}{d-3},-\frac{\ell}{d-3},1,1-\frac{M}{r^{d-3}}\right]\,.
\eea
At large values of $r$, the general solution of equation (\ref{ScalEqnSS}) can be expanded as
\bea\label{PhiAsymp}
\Phi=b_1 \left\{r^{\ell}\left[1+\OO(r^{-2})\right]+b_2 r^{3-d-\ell}\left[1+\OO(r^{-2})\right]\right\}\,.
\eea
The Love number is defined as $b_2$, but one encounters an ambiguity if $2\ell$ is divisible by $d-3$ since the $r^{3-d-\ell}$ term is present in both brackets. As described below, the resolution of this ambiguity is different for odd and even values of the ratio $\frac{2\ell}{d-3}$. If this ratio is not an integer, then $b_2$ is unambiguous, and the Love number is uniquely defined.

\bigskip
\noindent
Function (\ref{PhiHyper}) was used to compute Love numbers in \cite{StaticLN1,StaticLN2} leading to the following results:
\begin{enumerate}[(a)]
\item If integer $\ell$ is divisible by $(d-3)$, then the function in (\ref{PhiHyper}) is a polynomial of degree $\ell$ in $r$. This implies that $b_2=0$ in (\ref{PhiAsymp}), and the Love number vanishes.
\item If the ratio $k=\frac{\ell}{d-3}$ is neither an integer not half--integer, then using the relation
\bea\label{HyperAC}
F\left[-k,-k,1,1-z\right]=\frac{\Gamma[1+2k]}{\Gamma[1+k]^2}F[-k,-k,-2k,z]+
\frac{\Gamma[1-2p]}{\Gamma[1-p]^2}z^{2k+1}F[p,p,2p,z],\nonumber
\eea
where $p=k+1$, one can identify the first and the second term in the right--hand side with two brackets in (\ref{PhiAsymp}). This gives the parameters introduced in  (\ref{PhiAsymp}),
\bea
b_1=c_1\frac{\Gamma[1+2k]}{\Gamma[1+k]^2},\quad 
b_1 b_2=\frac{c_1}{b_1}\frac{\Gamma[-1-2k]}{\Gamma[-k]^2}M^{2k+1}\nonumber
\eea
and leads to the expression for the Love numbers 
\bea\label{LoveStaticAnsw}
\hskip -0.4cm
\la_\ell={b_2}=\frac{\Gamma[-1-2k]}{\Gamma[1+2k]}
\left[\frac{\Gamma[1+k]}{\Gamma[-k]}\right]^2M^{2k+1}=
\frac{(k+\frac{1}{2})\tan{\pi k}}{\pi}\frac{\Gamma[1+k]^4}{\Gamma[2(1+k)]^2}M^{2k+1}
\eea
We used the relation
\bea
\Gamma[q]\Gamma[-q]=-\frac{\pi}{q\sin(q\pi)}\,.
\eea
\item If the ratio $k=\frac{\ell}{d-3}$ is a half--integer, then to extract the asymptotic behavior of the wavefunction (\ref{PhiHyper}) at large values of $r$, one can use the relation
\bea\label{HyperAClog}
&&
\hskip -1cm
F[-k,-k,1,1-z]=\frac{\Gamma[2k+1]}{\Gamma[k+1]^2}\sum_{n=0}^{2k}\frac{[(-k)_n]^2}{n!(-2 k)_n}z^n\\
&&
\hskip -1cm
\qquad-\frac{(-1)^{2k}}{\Gamma[-k]^2}z^{2k+1}\sum_{n=0}^{\infty}\frac{(1+k)_n(1+k)_n}{n! (n+2k+1)!}z^{n}[\ln{z^{-1}}+
H_n+H_{n+2k+1}-2H_{n+k}].
\nonumber
\eea
Here $H(w)$ are the harmonic numbers defined by
\bea
H_w=\psi(w+1)+\gamma=\left.\frac{\d}{\d z}F[z,w,w,1]\right|_{z=0}\,.\nonumber
\eea
As we will demonstrate in the rest of this section, the terms containing $(\ln{z^{-1}})$ are fully determined by the expansion of the differential equation (\ref{ScalEqnSS}) near $z=0$, so they are not sensitive to the boundary conditions in the interior of the geometry and 
thus they carry no information about black holes.

Substitution of the expansion (\ref{HyperAClog}) into (\ref{PhiHyper}) gives
\bea\label{StaticRunOne}
\Phi=b_1\left\{r^\ell \left[1+\sum_{s=1}^{d+\ell-2}c_s r^{-s}\right]+
r^{3-d-\ell}\left[b_3 \ln\frac{M^{\frac{1}{d-3}}}{r}+b_2 +\OO(\frac{\ln r}{r})\right]\right\}
\eea
Then $b_2$ gives the Love number which carries information about the interior of the geometry, and $b_3$ determines the universal logarithmic running. The expansion (\ref{HyperAClog}) gives unambiguous expressions for $(b_2,b_3)$,
\bea\label{StaticKhalf}
&&b_{3}={{\hat b}_3}{M^{2k+1}}\,,\quad
b_2=b_3\left[H_n+H_{n+2k+1}-2H_{n+k}\right],\nn
&&{\hat b}_3=-\frac{(d-3)\Gamma[k+1]^2}{\Gamma[-k]^2(2k)! (2k+1)!}\,.
\eea
It is the parameter ${b}_2$ that should be interpreted as the Love number for the modes with half--integer $k$.

\end{enumerate}
For rotating black holes, the Love numbers are different, but unfortunately we were not able to compute them explicitly beyond five dimensions\footnote{For the five--dimensional black holes, the Love numbers were evaluated in \cite{RSST}. In the Appendix \ref{App5d} we review this construction and present an alternative parameterization of the solution which reduces to (\ref{PhiHyper}) once the rotation is turned off. Article \cite{RSST} treated rotating black holes in five dimensions and the general static solution (\ref{PhiHyper}) in all dimensions as disjoint cases.}.  For slow--rotating geometries, the corrections to (\ref{LoveStaticAnsw}) are small, while for option (a) the rotation constitutes the leading contribution. We will evaluate this contribution in section \ref{SecSlow}. 

\subsection{Logarithmic running for equal rotations in odd dimensions}
\label{SecLogSpec}

We begin the analysis of the coefficients specifying the logarithmic running by focusing on the special case of odd--dimensional black holes with equal rotation parameters. As discussed in section \ref{SecEqnSetup}, this leads to significant simplifications, and in this subsection we will focus on analyzing the solutions of the resulting radial differential equation (\ref{RadiaMPmainFin}). Introducing a new variable 
\bea\label{DefXra}
x=r^2+a^2,
\eea
we can rewrite (\ref{RadiaMPmainFin}) as
\bea\label{PhiXeqn}
4x^2(x^n-M x+Ma^2)\Phi''+4x(n x^n-M x)\Phi'-\ell(2n+\ell-2)x^n\Phi=0.
\eea
At large values of $x$ the general solution of this equation reads
\bea\label{PsiAsympLead}
\Phi=b_1 \left\{x^{\ell/2}\left[1+\OO(x)\right]+b_2 x^{-\ell/2-n+1}\left[1+\OO(x)\right]\right\}\,,
\eea
where $(b_1,b_2)$ are two integration constants. One of them is fixed by normalization, then the other one is determined by the regularity condition at the horizon. As discussed earlier, the Love number is defined as $b_2$. As reviewed in the previous subsection, for some values of $\ell$ the first bracket in (\ref{PsiAsympLead}) may contain terms which lead to logarithmic contributions to $\Phi$, and which should not be confused with the second term in (\ref{PsiAsympLead}). To isolate such contributions, we write (\ref{PsiAsympLead}) in more details as 
\bea\label{PsiAsymp}
\Phi&=&b_1 x^{\ell/2}\Bigg\{\left[1+\sum_{s=1}^{\ell+n-2}c_s x^{-s}\right]+
b_2 x^{-\ell-n+1}\phi_2+
b_3 x^{-\ell-n+1}\phi_3 \ln x \Bigg\}\,,\\
\phi_2&=&1+\sum_{s=1}^{\infty}{\tilde c}_s x^{-s},\quad 
\phi_3=1+\sum_{s=1}^{\infty}{\hat c}_s x^{-s}\,.\nonumber
\eea
To avoid unnecessary complications, we will set $b_1=1$. 
While the value of $b_2$ is determined by imposing the regularity condition at the horizon, the value of $b_3$ is fixed by the differential equation itself, and thus it is not sensitive to the interior. Therefore, it is misleading to call the parameter $b_3$ a ``running Love number''. Nevertheless, since this parameter has been extensively studied in the literature for static back holes, in this subsection we will determine it for the rotating case. The proper Love number, $b_3$, will be analyzed in subsequent sections. 

We begin with revisiting the static case. Setting $a=0$ in (\ref{PhiXeqn}) and introducing a new variable $z=x^{n-1}$, we find
\bea
z(z-M)\frac{d^2\Phi}{dz^2}+(2z-M)\frac{d\Phi}{dz}-\frac{\ell(2n+\ell-2)}{4(n-1)^2}\Phi=0.
\nonumber
\eea
This implies that the expansions in (\ref{PsiAsymp}) go in inverse powers of $z$ rather than $x$. 
Therefore, the logarithmic terns cannot be generated unless $x^{-\ell-n+1}$ can be written as an integer power of $z$, i.e., unless $\ell$ is divisible by $(n-1)$. As expected, this simple observation agrees with the results reviewed in subsection \ref{SecStatic}, but it also provides some insights on extension to the rotating case.  

Once rotation is turned on, all integer inverse powers of $x$ are present in the expansions (\ref{PsiAsymp}), but it is still convenient to parameterize $\ell$ in terms of two integers $(k,p)$ as
\bea\label{EllAsP}
\ell=k(n-1)+p,\quad p<n-1.
\eea 
Explicit analysis of the expansions (\ref{PsiAsymp}) indicates that at sufficiently large values of $n$ (the specific conditions on $n$ will be formulated below), function $\Phi$ can be written as 
\bea\label{PsiAsympBlock}
\Phi&=&x^{\ell/2}\Bigg\{\sum_{s=0}^{k+1}f_s(y) x^{-s(n-1)}+
x^{-(k+1)(n-1)-p}\left[b_2\phi_2+
b_3 \phi_3 \ln x\right] \Bigg\}\,,
\eea
where each $f_s(y)$ is a finite polynomial in $y=x^{-1}$, and the degree of this polynomial does not depend on $n$. Note that the normalization $b_1=1$ in (\ref{PsiAsymp}) translates into the requirement $f_0(0)=1$. The analysis of functions $f_s$ is performed in the Appendix \ref{AppRun}, and here we summarize the results.
\begin{enumerate}[(a)]
\label{PageAZ}
\item Function $f_s$ is a polynomial of degree $s$ in variable $y$,
\bea\label{SolnFsMain}
f_s(y)=\sum_{q=0}^s f_{sq}y^q\,,\quad s\le k\,.
\eea
All these functions are fully determined, for example,
\bea
f_1(y)=-\frac{M[k(n-1)+p]}{4(n-1)}+\frac{a^2M[k(n-1)+p][k(n-1)+p-2]}{4n[k(n-1)+p-1]}\,y
\eea
All other functions can be found as well, and the coefficients have the functional form
\bea
f_{sq}=a^{2q}M^s\,{\hat f}_{sq}\,.
\eea
The explicit expressions for numerical coefficients ${\hat f}_{sq}$, which do not depend on $(M,a)$, are not very illuminating. The conclusion about functions $f_s$ was reached assuming that all terms in  (\ref{PsiAsympBlock}) are independent, and this happens only at sufficiently large values of $n$. Indeed, the condition $s+1<n-1$ should be satisfied for all relevant values of $s$ in the first sum in (\ref{PsiAsympBlock}), leading to the restriction $n>k+1$. Smaller values of $n$ must be analyzed separately without relying on (\ref{PsiAsympBlock}), and this is done in the Appendix \ref{AppRun} as well, leading to the results presented in item (f). Items (b)--(e) focus on $n>k+1$.
\item For sufficiently large values of $n$, the expressions of $b_3$ are given by
\bea\label{B3pkMain}
k<n-1\ \Rightarrow\ b_3=B_3 a^{2p} M^{k+1}\,,
\eea
where $B_3$ are some numerical coefficients, which may or may not vanish. In particular, for $a=0$ we recover the static result (\ref{StaticKhalf}): there is no running for non--zero values of $p$.
\item In the static case, the running occurred only for $p=0$ and odd values of $k$, and equation (\ref{B3pkMain}) implies that this result is not modified by the rotation parameter: 
\bea\label{P0generic}
&&\hskip -1cm p=0,\ k=2{\hat k},\ n>k+1\quad \Rightarrow\quad b_3=0;\\
&&\hskip -1cm p=0,\ k=2{\hat k}+1,\ n>k+1\quad \Rightarrow\quad 
b_3=-\frac{(d-3)}{k! (k+1)!}
\frac{\Gamma[\frac{k}{2}+1]^2}{\Gamma[-\frac{k}{2}]^2}
{M^{k+1}}\,.\nonumber
\eea
\item Coefficient $b_3$ vanishes for sufficiently large values of $p$:
\bea\label{B3zeroPlar}
p> k+1\quad\Rightarrow\quad b_3=0\,.
\eea
In this case there is no logarithmic running even in the rotating case. 
\item The values of $B_3$ for $k=1,2,3$ are summarized in equations (\ref{RunK1gen}), (\ref{RunK2gen}), (\ref{RunK3gen}), and here we just quote the answers for $k=1$:
\bea\label{RunK1genMain}
k=1:\quad
p=1\ \Rightarrow\ B_3=-\frac{(n-2)n^2}{8(2n-1)(n-1)}\,,\ 
p=2\ \Rightarrow\ B_3=\frac{(n^2-1)^2}{32n^2}\,.
\eea
Interestingly, $B_3$ vanishes for $k=p=2$. 
\item For $n\le k+1$, the logic that led to equation (\ref{B3pkMain}) is no longer applicable, and a new analysis must be perfornmed. In particular, for $n=2$ one finds a simple modification of the static answer (here $k=2{\hat k}+1$):
\bea
n=2:&&
\ell=2{\hat k}+1\ \Rightarrow\ b_3=\frac{(d-3)\Gamma[\frac{k}{2}+1]^2}{\Gamma[-\frac{k}{2}]^2 k! (k+1)!}
[M(M-4a^2)]^{{\hat k}+1},\nn 
&&\ell\ne 2{\hat k}+1\ \Rightarrow\ b_3=0.
\eea
To see this, we observe that in terms of the coordinate ${\hat x}$ defined by (\ref{Xhyper1})--(\ref{Xhyper2}),
\bea\label{Xhyper1maim}
{\hat x}&=&\frac{1}{\sqrt{M(M-4a^2)}}\left[r^2+(\sqrt{M}-\sqrt{M-4a^2})^2\right]\nn
&=&
\frac{1}{\sqrt{M(M-4a^2)}}\left[x+(\sqrt{M}-\sqrt{M-4a^2})^2-a^2\right]
\eea
the unique regular solution of equation (\ref{PhiXeqn}) is given by (\ref{5dHyperNC})\footnote{See Appendix \ref{App5d} 
for detail.}:
\bea\label{5dHyperNCmain}
\Phi={\hat x}^{\frac{\ell}{2}} F[-\frac{\ell}{2}, -\frac{\ell}{2}, 1, 1 - \frac{1}{\hat x}]\,.
\eea
This implies that the Love number is given by the static answer (\ref{StaticKhalf}) with the replacement $M\rightarrow \sqrt{M(M-4a^2)}$.
Subsequent items will focus on $n\ge 3$. 
\item If $2<n\le k+1$, then a separate analysis that does not rely on equation (\ref{PsiAsympBlock}) is required, and it is carried out in the Appendix \ref{AppRun}. This leads to a modification of the relation (\ref{B3pkMain}): 
\bea\label{B3pkSpecMain}
k\ge n-1>1\ \Rightarrow\ b_3=B_3 a^{2p} M^{k+1}+{\hat B}_3 a^{2(p+n-1)} M^{k}\,,
\eea
The values of the numerical coefficients $(B_3,{\hat B}_3)$ for $k=1,2,3$ can be extracted from the relations (\ref{RunK1spec})--(\ref{RunK3spec}). Note that the value of $B_3$ in (\ref{B3pkSpecMain}) can be obtained by formally continuing equations (\ref{RunK1gen}), (\ref{RunK2gen}), (\ref{RunK3gen}) beyond the regime $n>k+1$ for which they were derived.
\item
To give some examples, we present the ${\hat B}_3$ corrections to the $p=0$ results (\ref{P0generic}):
\bea
&&\hskip -1.5cm (n,\ell)=(3,4)\ \Rightarrow\ b_3=\frac{2a^4 M^2}{27},\quad
(n,\ell)=(3,6)\ \Rightarrow\ b_3=\frac{9M^4}{2^{11}}\,,\\
&&\hskip -1.5cm (n,\ell)=(4,9)\ \Rightarrow\ b_3=\frac{27M^4}{2^{18}}[64-\frac{49a^6}{M}]\,,\
(n,\ell)=(3,8)\ \Rightarrow\ b_3=\frac{2\cdot 3^4\, (aM)^4}{5^3\cdot 7^2}\,,\nn
&&\hskip -1.5cm (n,\ell)=(4,12)\ \Rightarrow\ b_3=-\frac{24 a^6 M^4}{7\cdot 11^2}\,,
\quad (n,\ell)=(5,16)\ \Rightarrow\ b_3=\frac{28^2\, a^8M^4}{5^7}\,.\nonumber
\eea
Interestingly, ${\hat B}_3$ vanishes for $(n,\ell)=(3,6)$, so in this case the running coefficient $b_3$ is given by the static answer. This is the only case of ${\hat B}_3=0$ for $2<n<k+1<5$, the situations analyzed in the Appendix \ref{AppRun}. 
\end{enumerate}

\subsection{Logarithmic running for equal rotations in even dimensions}
\label{SecLogSpecEven}

Let us now analyze the logarithmic running for black holes in $d=2(n+1)$ dimensions assuming that all rotation parameters are equal. As demonstrated in section \ref{SecEqnSetup}, in this case the radial dynamics of the scalar field is governed by equation (\ref{RadialEqnEvenE}) with eigenvalues $(\la_1,\la_2)$ given by (\ref{LamEven}). We will now use the insights from the previous subsection to extract the running parameter from the solutions of (\ref{RadialEqnEvenE}), although the technical details will be radically different from the odd--dimensional case. 

\bigskip

Since equation (\ref{RadialEqnEvenE}) explicitly contains an odd power of $r$, transition to variable $x$ defined by (\ref{DefXra}) is nor convenient, so we will work directly in the $r$ coordinate. At large values of $r$, the solution of equation  (\ref{RadialEqnEvenE}) has the form\footnote{This is the counterpart of equation (\ref{PsiAsymp}) written in the $r$ rather than $x$ coordinate.}
\bea\label{PsiAsympEven}
\Phi&=&r^{\ell}\Bigg\{\left[1+\sum_{s=1}^{2(\ell+n-1)}c_s r^{-s}\right]+
r^{1-2\ell-2n}\left[b_2 \phi_2+b_3\phi_3 \ln r^2\right] \Bigg\}\,,\\
\phi_2&=&1+\sum_{s=1}^{\infty}{\tilde c}_s r^{-s},\quad 
\phi_3=1+\sum_{s=1}^{\infty}{\hat c}_s r^{-s}\,.\nonumber
\eea
As before, there are two constants: $b_2$ gives the Love number, and $b_3$ specifies the logarithmic running. To proceed, we write the parameter $\ell$ in terms of two integers $(K,p)$, mimicking the expression (\ref{EllAsP}):
\bea\label{EllAsPeven}
\ell=K(2n-1)+p.
\eea
Note that for $p=0$, equation (\ref{EllAsP}) gives $\ell=\frac{k(d-3)}{2}$, and (\ref{EllAsPeven}) gives $\ell=K(d-3)$, stressing the special role played by the values of $\ell$ divisible by $\frac{(d-3)}{2}$. The peculiarity of such values was already observed in the static case reviewed in section \ref{SecStatic}. 

Looking at equations (\ref{RadialEqnEvenE}) and (\ref{LamEven}),
\bea\label{RadEqnEvenCpy}
&&\frac{d}{dr}\left[\left\{(r^2+a^2)^n-M r\right\}\frac{d\Phi}{dr}\right]-(\la_1 r^2+\la_2 a^2)(r^2+a^2)^{n-2}\Phi=0,\\
&&\qquad\la_1=\ell (\ell+d-3),\quad  \la_2=(\ell+n-1)^2+\ell-q^2,\nonumber
\eea
we conclude that the odd powers of $r$ appear only in a specific combination 
\bea
\frac{M r}{\rho^{2n}},\quad \mbox{where}\quad \rho=\sqrt{r^2+a^2}.\nonumber
\eea
This suggests a refinement of the expansion (\ref{PsiAsympEven}):
\bea\label{PsiAsympEvenV2}
\Phi=r^\ell\Bigg\{\sum_{s=0}^{2K+1}\left[\frac{M r}{\rho^{2n}}\right]^s f_s(r)+
r^{1-2n-2\ell}\left[b_2\phi_2+
b_3 \phi_3 \ln r^2\right] \Bigg\}\,,
\eea
where functions $f_s$ do not contain the mass parameter. These functions are more complicated counterparts of finite polynomials $f_s$ appearing in (\ref{PsiAsympBlock}), but one can solve for them in the same way. For example, the leading terms is given by
\bea
f_0(r)=\frac{\Gamma[\ell+n]\Gamma[\ell+n-s]}{\Gamma[2(\ell+n)-1]}\frac{(2a)^{\ell+n-1}}{r^\ell \rho^{n-1}}i^{1-\ell-n-s}
P_{\ell+n-1}^{(s)} \left[\frac{ir}{a}\right]=1+\OO[(r/a)^2].
\eea
Here $P_{\ell+n-1}^{(s)}$ are associated Legendre polynomials. Similar explicit expressions can be found for other $f_s$, but they are not very illuminating. We will mention only one crucial property of $f_1$ which must be used to derive the results quoted below:
\bea
f_1(r)=-\frac{\ell}{2(2n-1)}+\OO(r^{-1}).\nonumber
\eea
The value of $b_3$ can be found by substituting expansions (\ref{PsiAsympEvenV2}) into the radial equation (\ref{RadEqnEvenCpy}) and performing analysis similar to the one presented in the Appendix \ref{AppRun}. For $n=1$, which corresponds to the Kerr black hole, one recovers the well--known result: there is no logarithmic running, and the Love numbers are trivial ($b_2=b_3=0$). Therefore, we focus on $n\ge 2$. In contrast to the odd-dimensional case, where there is no running for sufficiently large values of $p$ (recall equation (\ref{B3zeroPlar})), now $b_3$ is nontrivial for all nonzero $\ell$, as long as $n>1$. The general formula for the running parameter reads
\bea\label{B3pkEven}
b_3=\sum_{s=0}^K B^{(s)}_3 a^{2\ell-2s(d-3)} M^{2s+1}\,.
\eea
To compare this result with its odd--dimensional counterpart (\ref{B3pkSpecMain}), it is convenient to rewrite the latter relation as
\bea
d=2n+1:\qquad \ell=\frac{k}{2}(d-3)+p,\quad b_3=\sum_{s=k-1}^k {\hat B}^{(s)}_3 a^{2\ell-s(d-3)} M^{s+1}\,,\nonumber
\eea
We observe that in this expression the terms in the summations go in steps of $M$, in contrast to (\ref{B3pkEven}), where the steps are $M^2$. The powers of $a$ adjust accordingly. Furthermore, in contrast to the odd--dimensional case, where only ${\hat B}^{(k)}_3$ and ${\hat B}^{(k-1)}_3$ might be non--zero due to the relation (\ref{B3zeroPlar}) and its counterpart for $n\le k+1$, in even dimensions all $(K+1)$ terms are present in (\ref{B3pkEven}). 

The values of $B^{(s)}_3$ are computed using the logic outlined in the Appendix \ref{AppRun}, and here we present the results for small values of $\ell$. To do so, it is convenient to introduce two sets of constants:
\bea
\Lambda[P,Q]=(\la_2-P n-Q),\quad d_p=(2n+p).
\eea
For $\ell\le 4$ we find
\bea\label{OddLsumry}
&&\ell=0:\ b_3=0;\quad
\ell=1:\ b_3=\frac{\Lambda[2,0]\,a^2 M}{d_{-1}d_1};\quad
\ell=2:\ b_3=-\frac{\Lambda[4,2]\, \Lambda[0,2]\, a^4 M}{d_{-1}d_1 d_3}\,;\nn
&&\ell=3:\ b_3=\frac{(n+1)\,\Lambda[6,6]\, \Lambda[2,6]\, \Lambda[0,3]}{d_{-1}d_1(d_3)^2d_5}a^6 M\,;\\
&&\ell=4:\ b_3=-\frac{(n+1)\,\Lambda[8,12]\, \Lambda[4,12]\, \Lambda[2,9]\, \Lambda[0,4]}{
3d_{-1}d_1 d_3(d_5)^2d_7}a^8 M+\frac{16\,\Lambda[0,13]}{11\cdot 81\cdot 5}a^2 M^3\delta_{n.,2}\,.\nonumber
\eea
The first two lines in this summary correspond to $k=0$ in equation (\ref{EllAsPeven}) for all values of $n$ (we are assuming that $n>1$ since the four--dimensional case is well--known). The last line in (\ref{OddLsumry}) gives $k=0$ for $n>2$ and $k=1$ for $n=2$. This explains the second term in the latter case, in accordance with (\ref{B3pkEven}). The $n=2$ values is equally special in the $\ell=5$ case, where one finds
\bea\label{OddLsumryL5}
\ell=5:&& b_3=B^{(0)}_3 a^{10} M-\frac{32\,\Lambda[0,25]\, \Lambda[0,5]}{13\cdot 11\cdot 81\cdot 7}a^4 M^3\delta_{n,2}\,,\\
&&B^{(0)}_3=\frac{2\,\Lambda[10,20]\,\Lambda[6,20]\,\Lambda[4,17]\,\Lambda[2,12]\, \Lambda[0,5]}{
3 d_1 d_3(d_5 d_7)^2d_9}\,.\nonumber
\eea
In our final example, $\ell=6$, equation (\ref{B3pkEven}) gives $k=1$ for $n=\{2,3\}$, and the result reads
\bea
\ell=6:&& b_3=B^{(0)}_3 a^{12} M+{B}^{(1)}_3 a^{14-4n} M^3\,,\\
&&B^{(0)}_3=-\frac{2\,\Lambda[12,30]\,\Lambda[8,30]\,\Lambda[6,27]\,\Lambda[4,22]\,\Lambda[2,15]\, \Lambda[0,6]}{
15 d_1 d_3 d_5(d_7 d_9)^2d_{11}}\,,\nonumber\\
&&n=2\ \Rightarrow\ {B}^{(1)}_3=-\frac{64(\la_2-39)(\la_2-19)(97\la_2-738)}{3^2\cdot 5^5\cdot 77\cdot 13^2}\,,\nn
&&n=3\ \Rightarrow\ {B}^{(1)}_3=\frac{282-17\la_2}{5^3\cdot 7\cdot 17},\nn
&&n>3\ \Rightarrow\ {B}^{(1)}_3=0.
\nonumber
\eea
Our analysis has focused on $n>1$, but it is instructive to see how the four--dimensional results reviewed in the Appendix \ref{App4d} can be recovered from equation (\ref{RadEqnEvenCpy}). Applying the calculation performed in this subsection to the $n=1$ case, one finds a relation
\bea
b_3\propto (\la_2-\la_1).
\eea 
We now have to recall that in four dimensions there is only one separation constant, so in contrast to the general case, the parameters $\la_1$ and $\la_2$ in (\ref{RadEqnEvenCpy}) are not independent. This can be seen formally by recalling that $P_{n-1}$ defined by (\ref{PolynEven}) must be a polynomial of degree $(n-1)$, and for $n=1$ this happens if and only if $\la_2=\la_1$. This confirms that there is no logarithmic running in four dimensions, so one recovers a well-known result.

\subsection{Logarithmic running in the general case}
\label{SecLogGen}

The results of the last two subsections can be extended to Myers--Perry black holes with arbitrary rotation parameters. To do so, one should analyze the radial equations in the systems (\ref{OdeOddMPmain}) and (\ref{OdeEvenMPmain}) using the appropriate values of the quantized coefficients of polynomials $P_{n-1}$ and $P_{n-2}$. To treat the odd and even dimensions on the same footing, we 
will work in variable $r$ rather than a counterpart of $x$ defined by (\ref{DefXra}).

\bigskip
To study the logarithmic running produced by the radial equations (\ref{OdeOddMPmain}) and (\ref{OdeEvenMPmain}), it is convenient to write the rotation parameters as
\bea
a_s=a {\hat a}_s,\quad a^2=\frac{1}{n}\sum_{s=1}^{n} (a_s)^2\,.
\eea
Coefficients ${\hat a}_s$ are dimensionless, and all of them are equal to one in the special cases discussed in the last two subsections. As before, we write $\ell$ in terms of two integers $(k,p)$:
\bea\label{EllAsPall}
\ell=\frac{k}{2}(d-3)+p,\quad p<d-3,\quad \mbox{where}\ k(d-3)\ \mbox{is even}.
\eea
At large values of $r$, solutions of equations (\ref{OdeOddMPmain}) and (\ref{OdeEvenMPmain}) can be written in a form that generalizes (\ref{PsiAsympEvenV2}): 
\bea\label{PsiAsympGen}
\Phi=r^\ell\Bigg\{\sum_{s=0}^{k+1}\frac{M^s}{r^{s(d-3)}} f_s(r)+
r^{3-d-2\ell}\left[b_2\phi_2+
b_3 \phi_3 \ln x\right] \Bigg\}\,.
\eea
As in (\ref{PsiAsympEvenV2}), functions $f_s$ have no $M$ dependence, but now they contain both $a$ and dimensionless parameters ${\hat a}_s$. Normalization of $\Phi$ is fixed by the requirement $f_0(0)=1$. Repeating the arguments presented in the previous subsection, we arrive at a generalization of the expansion (\ref{B3pkEven}) for the running coefficient:
\bea\label{B3pkAll}
\mbox{even}\ d:&&b_3=\sum_{s=0}^{k/2} {\hat B}^{(2s)}_3 a^{2\ell-2s(d-3)} M^{2s+1}\,,\nn
\mbox{odd}\ d:&&b_3=\sum_{s=0}^{k} {\hat B}^{(s)}_3 a^{2\ell-s(d-3)} M^{s+1}\,.
\eea
Here ${\hat B}^{(q)}_3$ are homogeneous polynomials of degree $(2\ell-q(d-3))$ in dimensionless rotation parameters 
$\{{\hat a}_1,\dots {\hat a}_n\}$, and in even dimensions index $q$ takes only even values. The discussion of section \ref{SecStatic} implies that 
\bea
\mbox{even}\quad q=\frac{2\ell}{d-3}:&&{\hat B}^{(q)}_3=0,\nn
\mbox{odd}\quad q=\frac{2\ell}{d-3}:&&{\hat B}^{(q)}_3=-\frac{(d-3)\Gamma[1+\frac{q}{2}]^2}{\Gamma[-\frac{q}{2}]^2 q! (q+1)!}\,.
\eea
In particular, this implies that $b_3=0$ for $\ell=0$. 
Direct calculations show that ${\hat B}^{(q)}_3$ are non--zero for all other values of $q$, unless parameters 
$\{{\hat a}_1,\dots {\hat a}_n\}$ take special values, like the ones encountered in section \ref{SecLogSpec}, where all these numbers were equal. Therefore, generically, there is a nontrivial logarithmic running for all non--zero values of $\ell$.

\section{Love numbers for slowly rotating black holes}
\label{SecSlow}

In contrast to the logarithmic running, which can be deduced purely from solving the wave equation at large values of the radial coordinate, the Love numbers are sensitive to the interior of the geometry, and in the case of black holes, information about the interior is encoded in the boundary conditions for the wave function at the horizon. As demonstrated in the previous section, by looking at the wave equation far away from all sources, one uniquely determines the value of parameter $b_3$ in (\ref{PsiAsympEven}), but cannot deduce any conclusions about the value of $b_2$, which contains a free integration constant. In the static case reviewed in section \ref{SecStatic}, the value of $b_2$ was fixed by finding the exact solution of the wave equation everywhere, but unfortunately for rotating black holes such exact solution is not available beyond five dimensions\footnote{Interestingly, an exact solution can be found for extremal geometries in seven dimensions, and this case will be discussed in section \ref{SecExtremal}}. In the absence of exact solution, one has to rely on various approximations, and in this article we will focus on two of them: the WKB expansion, which is discussed in section \ref{SecWKB} and pertubation theory in the rotation parameter, which will be studied here. 

As reviewed in section \ref{SecEqnSetup}, a black hole in $d$ dimensions can have at most 
$[(d-1)/2]$ rotation parameters $a_k$, so to introduce the concept of ``slow rotation'' we write them as $a_k=a \beta_k$ with dimensionless parameters $\beta_k$ whose absolute value cannot exceed one. Then we will assume that $a\ll r_h$, where $r_h$ is the horizon radius of the black hole and expand the Love numbers in powers of $(a/r_h)$. Similar expansions can be performed in the differential equations as well, as long as $(r-r_h)\gg a$. We will begin with a detailed study of the case of equal rotation parameters in odd dimensions, for which the radial equation has a very simple form (\ref{RadiaMPmainFin}), and in section \ref{SecPertGen} we will extend these results to geometries with arbitrary rotations. Interestingly, the Love numbers are still given by the expressions derived in the case of equal rotations\footnote{To cover the even--dimensional case, one should formally extend these results by analytically continuing to even values of  $d$.}, as long as one introduces a single effective rotation parameter discussed on page \pageref{PageRotParam}.

\subsection{Geometries with equal rotations in odd dimensions}
\label{SecPertSpec}
In this subsection we consider black holes in $(2n+1)$ dimensions with $n$ equal rotation parameters, $a_k=a$, and we will use equation (\ref{PhiXeqn}) to evaluate the expansion of the Love numbers in the powers of $(a/r_M)$, where $r_M$ is the length scale associated with the mass\footnote{Specifically, $r_M=M^{\frac{1}{d-3}}$.}. Alternatively, one could look at the expansion in powers of $(a/r_h)$, where $r_h$ is the radius of the horizon, and the two approaches are related by a simple resummation of the series. Our final result will be given by (\ref{LovePertEqA}) with parameters $A_{n,k}$ summarized in (\ref{PertAvalues}).

Let us analyze the solutions of the radial equation equation (\ref{PhiXeqn}) :
\bea\label{PhiXeqnA}
4x^2\frac{d}{dx}\left[(x^n-M x+Ma^2)\frac{d\Phi}{dx}\right]-\ell(2n+\ell-2)x^n\Phi=0.
\eea
To avoid unnecessary clutter in subsequent formulas, we rescale $x$ and $a$ as
\bea
x\rightarrow x M^{\frac{1}{n-1}}\,,\quad a\rightarrow a M^{\frac{1}{2(n-1)}}\,.\nonumber
\eea
This amounts to setting $M=1$ in equation (\ref{PhiXeqnA}).

As discussed earlier, at large values of $x$ the general solution of equation (\ref{PhiXeqnA}) has the form (\ref{PsiAsympLead}), and the Love number $\lambda_\ell$ is given by the value of $b_2$. For slowly rotating black holes, one can expand the love number in power series as
\bea\label{LambExpandAA}
\lambda_\ell(a)=\lambda^{(0)}_\ell+a^2 \lambda^{(1)}_\ell+a^4 \lambda^{(2)}_\ell+\dots
\eea
Odd powers of $a$ are absent since the geometry (\ref{MPodd}) is symmetric under simultaneous change of sign in $(t,a)$. In the absence of rotation, the Love numbers were evaluated in \cite{StaticLN2}, and we reviewed these results in section \ref{SecStatic}\footnote{In the case of nontrivial logarithmic running, our results were new, as discussed in section \ref{SecStatic}.}. Let us now compute $\lambda^{(1)}_\ell$, the first correction caused by rotation. To do that, we expand the wavefunction in powers $a$:
\bea\label{PsiExpand}
\Phi=\Phi_0+a^2\Phi_1+\OO(a^4).
\eea
Substituting this expansion into the leading order of equation (\ref{PhiXeqnA}) and imposing regularity at the horizon, we find\footnote{Recall that we have set $M=1$ in equation (\ref{PhiXeqnA}). For other values of $M$, the argument of the hypergeometric function is 
$(1-M x^{1-n})$.}
\bea\label{Phi0}
\Phi_0=c_1 x^{\ell/2}F\left[-\frac{\ell}{2(n-1)},-\frac{\ell}{2(n-1)},1,1-x^{1-n}\right]\,.
\eea
As discussed in section \ref{SecStatic}, this expressions leads to non--trivial Love numbers unless the ratio $\frac{\ell}{2(n-1)}=\frac{\ell}{d-3}$ is an integer, so in this latter case, the leading contribution to the Love number is proportional to a positive power of $a$. Evaluation of such contributions is the main goal of this subsection.

\bigskip

Let us consider equation (\ref{PhiXeqnA}) with 
\bea\label{LasK}
\ell=2(n-1)k,
\eea
where $k$ is an integer. In this case, the leading solution (\ref{Phi0}) becomes a polynomial of degree $k$ in variable $x^2$. To get some intuition about the structure of $\Phi_1$, it is instructive to look at the first few values of $n$.
\begin{itemize}
\item For $n=2$, one finds that the general solution of equation for $\Phi_1$ has the form
\bea\label{Psi1anstzN2}
\Phi_1(x)=P_{k-1}(x)-\Phi_0(x)\left[e_1+e_2 \ln(1-x^{-2})\right],
\eea
Here $(e_1,e_2)$ are integration constants, and $P_{k-1}(x)$ is a polynomial of degree $(k-1)$ in $x$ whose coefficients depend on $(e_1,e_2)$. Parameter $e_2$ describes the counterpart of the solution (\ref{Phi0}) which is singular at the horizon, so one must set $e_2=0$. Parameter $e_1$ produces an $a$--dependent shift in the constant $c_1$, once the solution is substituted into (\ref{PsiExpand}), so without loss of generality we will set $e_1=0$ as well. Then $\Phi_1$ reduces to the polynomial $P_{k-1}(x)$. To find this polynomial, we observe that the exact solution for all values of $a$ is given by
\bea\label{5dHyperNCmain2}
\Phi={\hat x}^{\frac{\ell}{2}} F[-\frac{\ell}{2}, -\frac{\ell}{2}, 1, 1 - \frac{1}{\hat x}]\,,\quad
{\hat x}=
\frac{x+(\sqrt{M}-\sqrt{M-4a^2})^2-a^2}{\sqrt{M(M-4a^2)}}\,.
\eea
The details are presented on the Appendix \ref{App5d} (see also discussion leading to equations (\ref{Xhyper1maim}) and (\ref{5dHyperNCmain})). For integer values of $k=\ell/2$, function $\Phi$ is a polynomial of degree $k$ in ${x}$, which implies that 
$\Phi_0$ and $\Phi_1$ have degrees $k$ and $k-1$.
\item For $n=3$ one finds that the general solution of equation for $\Phi_1$ has the form
\bea\label{Psi1anstzN3}
\Phi_1(x)=P_{2k-1}(x)-\Phi_0(x)\left[c(k)\ln(1+x^{-1})+e_1+e_2 \ln(1-x^{-2})\right],
\eea
Here $P_{2k-1}(x)$ is a polynomial of degree $(2k-1)$ in $x$, $c(k)$ is a specific numerical coefficient, and $(e_1,e_2)$ are integration constants. As before, parameter $e_2$ should be set to zero for regularity at the horizon, and  parameter $e_1$ can be set to zero to avoid an $a$--dependent shift in the constant $c_1$. Thus we arrive at the final form of $\Phi_1$:
\bea\label{Psi1anstzN3fin}
\Phi_1(x)=P_{2k-1}(x)-c(k)\Phi_0(x)\ln(1+x^{-1}).
\eea
Parameter $c(k)$ can be evaluated on the case--by--case basis, for example,
\bea\label{CvaluesK3}
c(1)=\frac{4}{3}\,,\quad c(2)=\frac{108}{35}\,,\quad c(3)=\frac{56}{11}\,,\quad
c(4)=\frac{5960}{819}\,,\quad c(5)=\frac{443620}{46189}\,.
\eea
Recalling the asymptotic expression (\ref{PsiAsympLead}), we conclude that the polynomial $P_{2k-1}(x)$ does not contribute to $c_2$, so the Love number is completely determined by $c(k)$ and the hypergeometric polynomial $\Phi_0$. We will discuss the extraction of Love numbers from this data below, when we look at arbitrary values of $n$.
\item For $n=4$ one finds that the general solution of equation for $\Phi_1$ has the form
\bea\label{Psi1anstzN4}
\Phi_1(x)&=&P_{3k-1}(x)-\Phi_0(x)\left[c(k)F(x)+e_1+e_2 \ln(1-x^{-3})\right],\\
F(x)&=&\frac{1}{2}\left\{\ln\left[\frac{1-x^3}{x^2(1-x)}\right]+\frac{i}{\sqrt{3}}
\ln\left[\frac{x+\frac{1}{2}-\frac{i\sqrt{3}}{2}}{x+\frac{1}{2}+\frac{i\sqrt{3}}{2}}\right]\right\}\,.\nonumber
\eea
As before, regularity condition implies that $e_2=0$ and rescaling of function $\Phi$ can be used to set $e_1=0$. Therefore, the final expression for $\Phi_1(x)$ reads
\bea\label{Psi1anstzN4a}
\Phi_1(x)&=&P_{3k-1}(x)-c(k)\Phi_0(x)F(x).
\eea
The values of $c(k)$ can be determined on the case--by--case basis, as in (\ref{CvaluesK3}). While function $F$ is rather complicated, its derivative is remarkably simple:
\bea\label{DerJn4}
\frac{dF}{dx}=-\frac{1-x^2}{x(1-x^3)}\,.
\eea
The coefficient $c(k)$ determines the Love numbers, as we will demonstrate below for a more general case of arbitrary $n$.
\end{itemize}
To construct the function $\Phi_1(x)$ for arbitrary values of $n$, we will use the $n=2,3,4$ cases discussed above as inspirations for the form of the solution. Specifically, observing that the derivative of the logarithm appearing in (\ref{Psi1anstzN3}),
\bea
\frac{d}{dx}\ln(1+x^{-1})=-\frac{1-x}{x(1-x^2)}\,,
\eea
looks similar to (\ref{DerJn4}), we define a function $J_n$ by
\bea
J_n(x)=-\int \frac{1-x^{n-2}}{x(1-x^{n-1})}dx
\eea
Then inspired by (\ref{Psi1anstzN2}), (\ref{Psi1anstzN3fin}) and (\ref{Psi1anstzN4a}), we write the solution for $\Phi_1$ regular at $x=1$ in the form
\bea\label{Psi01}
\Phi_1(x)=P_{(n-1)k-1}(x)-\Phi_0(x)\left[A_{n,k} J_n+e_1\right],
\eea
where $e_1$ is an undetermined constant, and polynomial $P_{(n-1)k-1}$ is uniquely fixed. 

Direct calculations show that
\bea\label{PertAvaluesPre}
&&A_{n,1}=\frac{2(n-1)}{2n-3},\quad 
A_{n,2}=A_{n,1}\left[3-\frac{12(n-1)(n-2)}{(3n-4)(4n-5)}\right],\nn
&&A_{n,3}=A_{n,1}\left[6-\frac{60(n-1)(n-2)(24-41n+18 n^2)}{(3n-4)(4n-5)(5n-6)(6n-7)}\right]
\\
&&A_{n,4}=A_{n,1}\left[10-\frac{60(n-1)(n-2)(3360 - 11696 n + 15373 n^2 - 9050 n^3 + 2016 n^4)}{(3n-4)(4n-5)(5n-6)(6n-7)(7n-8)(8n-9)}\right]\nonumber
\eea
The large numbers involved in these expressions are not very illuminating, so it might be useful to observe that somewhat nicer structures arise if the coefficients $A_{n,k}$ are written in terms of parameters $b_s=s(n-1)-1$, which determine the structure of poles in (\ref{PertAvaluesPre}). Isolating such poles, one finds: 
\bea\label{PertAvalues}
A_{n,2}&=&A_{n,1}\left[2+\frac{8}{b_3}-\frac{9}{b_4}\right]\,,\nn
A_{n,3}&=&A_{n,1}\left[3+\frac{5\cdot 2^4}{b_3}-\frac{5\cdot 7\cdot 3^2}{b_4}+
\frac{2^4 3^3}{b_5}-\frac{2^3 5^2}{b_6}\right]\,,\\
A_{n,4}&=&A_{n,1}\left[4+\frac{2^4 5^2}{b_3}-\frac{3\cdot 41\cdot 5^2}{b_4}+
\frac{21\cdot 29\cdot 2^4}{b_5}-\frac{77\cdot 2^3 5^2}{b_6}+
\frac{3\cdot 2^5 5^3}{b_7}-\frac{3\cdot 7^2 5^2}{b_8}\right]\,.\nonumber
\eea
To extract the Love numbers, we recall that function $\Phi_0$ given by (\ref{Phi0}) can be rewritten as 
\bea
\Phi_0=c_1 x^{-k(n-1)}F\left[-k,-k,1,1-x^{n-1}\right]\,,
\eea
and the Love number is  determined for the coefficient in front of $x^{n-1+(n-1)k}=x^{(n-1)(k+1)}$ in $\Phi$. This gives the leading $a$--dependent contribution to the Love numbers (\ref{LambExpandAA}) in terms of constants $A_{n,k}$:
\bea\label{LovePertEqA}
\lambda^{(1)}_{n,k}=-A_{n,k}\frac{2}{n-1}\frac{k!}{2^{k+1}(2k+1)!!}\,,\quad k=\frac{\ell}{2(n-1)}\,.
\eea
For completeness we also recalled the expression for $k$ from (\ref{LasK}), and since in this subsection we focused on integer values of $k$, Love numbers vanish for $a=0$: $\lambda^{(0)}_{n,k}=0$.

\subsection{General Myers--Perry black holes}
\label{SecPertGen}

Let us now generalize the results of the previous subsection to black holes with arbitrary rotation parameters $(a_1,\dots a_n)$. Since we are focusing on slowly rotating geometries, we will assume all $a_k$ are small in comparison to the length scale set by mass, and to make this assumption more explicit, we write the rotation parameters in terms of a single scale $b$ and dimensionless numbers $\beta_k$:
\bea
a_k=\beta_k b,\quad |\beta_k| <1.
\eea
We will assume that at least one of the parameters $\beta_k$ is of order one, so $b$ is the characteristic scale of the rotation. Expanding the radial equations (\ref{OdeOddMPmain}), (\ref{OdeEvenMPmain}) up to second power in $b$, we find
\bea\label{EqnAbrAeven}
r\frac{d}{dr}\left[\frac{r^{d-1}-M r^2+w_1 b^2 r^{d-3}}{r}\dot\Phi\right]-
\Lambda r^{d-3}\left[1+\frac{w_2b^2}{r^2}\right]\Phi+\OO(b^4)=0.
\eea 
Here $\Lambda=\ell(\ell+d-3)$, and $(w_1,w_2)$ are dimensionless quantities depending on parameters $\beta_k$ and coefficients of the polynomials appearing in (\ref{OdeOddMPmain}), (\ref{OdeEvenMPmain}). As in the previous subsection, we write
\bea
\Phi=\Phi_0+b^2\Phi_1+\OO(b^4),
\eea
and as discussed in section \ref{SecStatic}, the leading contribution to this solution is
\bea
\Phi_0=c_1 r^{\ell}F\left[-\frac{\ell}{d-3},-\frac{\ell}{d-3},1,1-r^{3-d}\right]\,.
\eea
The equation for $\Phi_1$ reads
\bea\label{EqnPhi1GenA}
r\frac{d}{dr}\left[\frac{r^{d-1}-M r^2}{r}\dot\Phi_1\right]-
\Lambda r^{d-3}\Phi_1=-r\frac{d}{dr}\left[\frac{w_1 r^{d-3}}{r}\dot\Phi_0\right]+
\Lambda r^{d-3}\frac{w_2}{r^2}\Phi_0.
\eea 
In particular, the special case discussed in the previous subsection had
\bea
w_1=n,\quad w_2=n-2,\quad {n}= \frac{d-3}{2}+1\,.
\eea
We will now argue that the Love numbers corresponding to the general equation (\ref{EqnPhi1GenA}) can be reduced to the results for that special case. 
\begin{itemize}
\item
We begin with observing that if the parameters in (\ref{EqnPhi1GenA}) are related by
\bea\label{w1w2Bk}
w_1=B_{n,k}w_2,
\eea
with constants $B_{n,k}$ given below in (\ref{BnkNimbers}), then equation (\ref{EqnPhi1GenA}) has a polynomial solution in the form
\bea
\Phi^{(polyn)}_1=\sum_{s=1}^k u_s r^{2s(d-3)-2}\,,
\eea
and this contribution to $\Phi$ does not affect the Love numbers. 
\item
Writing parameters $(w_1,w_2)$ as\footnote{Note the parameter $n$ is either integer or half--integer depending on the dimension, but formally the perturbative discussion from the last subsection applied to both cases.}
\bea\label{w1w2Rotat}
w_1=n{\bar w}+B_{n,k}w_3,\quad w_2=(n-2){\bar w}+w_3,\quad {n}= \frac{d-3}{2}+1,
\eea
we conclude that the most general solution of equation (\ref{EqnPhi1GenA}) has the form
\bea
\Phi_1={\bar \Phi}_1+w_3\Phi^{(polyn)}_1,
\eea
where ${\bar \Phi}_1$ satisfies the equation
\bea
r\frac{d}{dr}\left[\frac{r^{2n}-M r^2}{r}\dot{\bar\Phi}_1\right]-
\Lambda r^{2n-2}{\bar\Phi}_1=-r\frac{d}{dr}\left[\frac{n{\bar w} r^{2n-2}}{r}\dot\Phi_0\right]+
\Lambda r^{2n-2}\frac{(n-2){\bar w}}{r^2}\Phi_0.\nonumber
\eea 
This implies that the function
\bea
{\bar\Phi}\equiv \Phi_0+b^2 {\bar\Phi}_1,
\eea
which leads to the same Love numbers as $\Phi$, obeys the truncated version of equation (\ref{RadiaMPmainFin}) encountered for equal rotations
\bea\label{RadGenerA2}
r\frac{d}{dr}\left[\frac{(r^2+a^2)^n-2M r^2}{r}\frac{d{\bar\Phi}}{dr}\right]-\la (r^2+a^2)^{n-2}{\bar\Phi}+\OO(a^4)=0\,.
\eea
\label{PageRotParam}
Here we defined a new effective rotation parameter $a$ through the relation $a^2={\bar w}b^2$.
\item As demonstrated in the previous subsection, the unique solution of equation (\ref{RadGenerA2}) regular at the horizon reads
\bea\label{PhiBarOne}
{\bar\Phi}(x)=\Phi_0+a^2\left\{P(x)-\Phi_0\left[A_{n,k} J_n+e_1\right]\right\}+\OO(a^4)
\eea
Then the Love numbers are given by equation  (\ref{LovePertEqA}): 
\bea\label{LoveGenRot}
\lambda_\ell(a)=a^2 \lambda^{(1)}_{n,k}+\OO(a^4)=-a^2\, \frac{2A_{n,k}}{n-1}\frac{k!}{2^{k+1}(2k+1)!!}
+\OO(a^4).
\eea
The relevant solution of the original equation (\ref{EqnAbrAeven}) is
\bea\label{PhiBarTwo}
{\Phi}(x)={\bar\Phi}(x)+w_3\Phi^{(polyn)}_1\,,
\eea
with $w_3$ defined through (\ref{w1w2Rotat}). 
\item
Parameters $B_{n,k}$ appearing in the relation (\ref{w1w2Bk}) can be calculated for all values of $(n,k)$, but the explicit expressions for them are not very illuminating. For completeness we present the first few families of $B_{n,k}$ corresponding to $k=1,2,3,4$:
\bea\label{BnkNimbers}
B_{n,k}&=&\frac{1}{p}(p-1)^2(2p-1)\left[2p^2-p+1+p(p+1){\overline{B}}_{n,k}\right],\quad
p\equiv \frac{d-3}{2},\nn
{\overline{B}}_{n,1}&=&0,\quad {\overline{B}}_{n,2}=\frac{4}{12p^2-5p+1},
\quad {\overline{B}}_{n,3}=\frac{10(18p^2-p+1)}{360p^4-258p^3+89p^2-12p+1}\,,\nn
{\overline{B}}_{n,4}&=&\frac{18(672p^4-82p^3+83 p^2-2p+1)}{
20160p^6-19128p^5+8770p^4-2019p^3+301p^2-21p+1}
\eea
\end{itemize}
To summarize, we have demonstrated that equation (\ref{EqnAbrAeven}) for a scalar field in a slowly rotating Myers--Perry geometry with arbitrary but small rotation parameters has the unique solution (\ref{PhiBarOne}), (\ref{PhiBarTwo}). The resulting Love numbers are given by (\ref{LoveGenRot}), where $a^2={\bar w}b^2$, and ${\bar w}$ is defined through (\ref{w1w2Rotat}).

\section{Love numbers for extremal black holes}
\label{SecExtremal}

So far we have been focusing on non--extrenal black holes, but calculation of the Love numbers in the extremal case requires a special discussion since the boundary conditions at the horizon are modified. We will consider this case here focusing on equal rotations in odd dimensions. 

Let us begin with discussing the extremal limit in equation (\ref{RadiaMPmainFin}). The parameters of this limit are defined by two conditions:
\bea
\Delta(r_h)=0,\quad \Delta'(r_h)=0,\quad\mbox{where}\quad
\Delta(r)=(r^2+a^2)^n-M r^2.
\eea
This gives
\bea
r_h^2=\frac{a^2}{n-1},\quad M=n\left[\frac{n}{n-1}\right]^{n-1}a^{2(n-1)}\equiv
\mu_n a^{2(n-1)}
\eea
Since parameters $M$ and $a$ are now related the perturbative approach discussed in section \ref{SecSlow} is no longer appropriate, and one has to rely on the exact solution. Fortunately, such solutions can be obtained in d=5 and d=7. As in section \ref{SecLogSpec} we introduce a shifted radial variable, but since now the rotation parameter represents the only scale in the problem, we will make this coordinate dimensionless:
\bea
x=\frac{r^2+a^2}{a^2}
\eea 
Then the final equation becomes a modified version of (\ref{PhiXeqnA}):
\bea\label{PhiXeqnAextr}
4x^2\frac{d}{dx}\left[[x^n-\mu_n(x-1)]\frac{d\Phi}{dx}\right]-\ell(2n+\ell-2)x^n\Phi=0.
\eea
Let us now discuss solutions of this equation in five and seven dimensions.

\bigskip
\noindent
{\bf Five dimensions}

In five dimensions, equation (\ref{PhiXeqnAextr}) becomes
\bea
4\frac{d}{dx}\left[(x-2)^2\frac{d\Phi}{dx}\right]-\ell(\ell+2)\Phi=0.
\eea
The most general solution is given by 
\bea\label{5dExtrSoln}
\Phi=c_1 (x-2)^{\ell/2}+c_2(x-2)^{-1-\ell/2}\,,
\eea
and regularity at the horizon requires $c_2=0$. As expected, at infinity the solution behaves as $r^\ell$. The Love number is defined as the coefficient in from of $r^{-\ell-1}$, and it is given by
\bea
\lambda_{\ell} = \frac{2^{\ell+1} a^{2(\ell+1)} \Gamma[\frac{\ell}{2}+1]}{
\Gamma[-\frac{\ell}{2}] \Gamma[\ell+2]}
\eea
This number vanishes for even $\ell$, which can also be seen directly from (\ref{5dExtrSoln}) since in this case the solution is a polynomial which has no negative powers of $r$. 

\bigskip
\noindent
{\bf Seven dimensions}

In seven dimensions,  equation (\ref{PhiXeqnAextr}) becomes
\bea\label{Extrem7deqn}
\frac{d}{dx}\left[(x+3)(2x-3)^2\frac{d\Phi}{dx}\right]-\ell(\ell+4)x\Phi=0.
\eea
The solution of this equation regular at the horizon reads
\bea\label{Extrem7dsoln}
\Phi=z^{(s-1)/2}
(x+{3 })^{-(\ell+4)/2}  
 F\left[\alpha, \alpha, s+1, z\right] 
\eea
Here we defined a convenient variable $z$, as well as two parameters, $\alpha$ and $s$:
\bea
z=\frac{2x - 3 }{2 (x + 3)},\quad \alpha=\frac{\ell+s+3}{2},
\quad s=\sqrt{\frac{(\ell+1)(\ell+3)}{3}}
\eea
The second solution of equation (\ref{Extrem7deqn}) diverges near the horizon as $z^{-(s+1)/2}$ and thus should be discarded. 
At large values of $x$, function $\Phi$ can be expanded as (\ref{PsiAsymp}). The relevant terms have the form  
\bea\label{PsiAsympExtrm}
\Phi&=&b_1 x^{\ell/2}\Bigg\{1+b_2 x^{-\ell-2}+
b_3 x^{-\ell-2} \ln x+\dots \Bigg\}\,,
\eea
where $b_2$ gives the Love number, and $b_3$ describes the logarithmic running. Direct evaluation gives
\bea\label{PsiRunExtrm}
b_2&=&b_3\left\{\log\left[\frac{9}{2}\right]+2H\left[\frac{\ell+s+1}{2}\right]\right\}+
{\bar b}_2+{\hat b}_2 s\\
b_3&=&-\frac{(-9/2)^{\ell+2}}{(\ell+1)!(\ell+2)!}
\left[\frac{\Gamma[(\ell+s+3)/2]}{\Gamma[(s-\ell-1)/2]}\right]^2\nonumber
\eea
Here $({\bar b}_2,{\hat b}_2)$ are rational numbers depending on $\ell$, and $H(w)$ are the harmonic numbers defined by
\bea
H[w]=\psi(w+1)+\gamma=\left.\frac{\d}{\d z}F[z,w,w,1]\right|_{z=0}\,.
\eea
The explicit expressions for $({\bar b}_2,{\hat b}_2)$ are not very illuminating, so to avoid unnecessary clutter in this discussion, we present then in the Appendix \ref{AppExtremal}. 

It is interesting to look at the special values of $\ell$ which avoid the logarithmic running in (\ref{PsiAsympExtrm}). This happens when the gamma function in the denominator of $b_3$ goes to infinity, i.e., when 
\bea\label{PellMain}
s-\ell-1=-2k
\eea
with a non---negative integer $k$. Remarkably, this equation admits an infinite set of solutions for integer values of $(\ell,k)$. This problem is analyzed in the Appendix \ref{AppExtremal}, and here we just quote the answer:
\bea\label{EllFromPell}
\ell= \frac{1}{2} \left[(2 - \sqrt{3})^p + (2 + \sqrt{3})^p\right]- 2\,,
\eea
where $p$ is a positive integer. The first few values are\footnote{Note that $p=0$ gives an unphysical value $\ell=-1$.}
\bea
\ell=0,\,5,\,24,\,95,\,360.
\eea

\bigskip
\noindent
{\bf Nine dimensions}

While we were not able to construct explicit solutions of equation (\ref{PhiXeqnAextr}) beyond seven dimensions for arbitrary values of angular momentum $\ell$, interestingly, in nine dimensions one one gets a hypergeometric equation for the special unphysical values $\ell={-2,-4}$. The arguments isolating the specal role of these numbers are presnted at the end of the Appendix \ref{AppExtremal}, and here we just start from the resulting equation (\ref{PhiXeqnAextr}):
\bea\label{PhiXeqnAextr9d}
\frac{d}{dx}\left[[3x-4]^2[3x^2+8x+16]\frac{d\Phi}{dx}\right]+54x^2\Phi=0.
\eea
Two linearly independent solutions of this equation can be chosen to be
\bea\label{9dsolnHyper}
\Phi_\pm=\left[\frac{3x-4}{x-x_1}\right]^{\frac{s-1}{2}}\hspace{-0.3cm}
 \frac{1}{x-x_1}F\left[\frac{s+1}{2},\frac{s+1}{2},s+1,\frac{\mu(3x-4)}{3(x-x_1)}\right],\quad s=\pm \frac{i}{\sqrt{3}}
\eea
Here
\bea
\mu=\frac{2(1+i\sqrt{2})}{3},\quad x_1=-\frac{4(1-i\sqrt{2})}{3}
\eea
As in seven dimensions, the Love numbers can be extracted by expanding the solution at infinity, but the results are not very interesting since the angular solutions corresponding to $\ell={-2,-4}$ are not normalizable\footnote{The ``eigenvalue'' $\ell(\ell+6)=-8$ entering equation (\ref{PhiXeqnAextr9d}) is negative, so it cannot correspond to normalizable modes on a compact angular space.}, and thus these modes are not physical.

\section{Love numbers in the WKB approximation}
\label{SecWKB}

In section \ref{SecSlow} we computed Love numbers for small values of rotation parameters, but unfortunately exact evaluation of these numbers in six and higher dimensions is beyond 
reach, apart from some extremal cases discussed in section \ref{SecExtremal}. As expected the problem simplifies when some parameters become large. For example, the authors of \cite{LoveRotHd1} computed the Love numbers for large spacetime dimensions. In this section we will use a different approximation and compute the Love number for the large values of angular momentum using the leading contribution to the WKB expansion.  

\bigskip

Let us consider equation (\ref{PhiXeqnA}) for large angular momentum $\ell$. 
In the WKB approximation the solution becomes
\bea\label{PhiWKB}
\Phi=c_1 e^{(\ell+n-1) S_+}+c_2 e^{-(\ell+n-1) S_-}\,,
\eea
where functions $S_\pm$ have expansions
\bea
S_\pm=S_0\pm \xi S_1 +\xi^2 S_2\pm \xi^3 S_3+\dots,\quad \xi=\frac{1}{\ell+n-1}.
\eea
We will focus only on the first two terms in the WKB expansion, $S_0$ and $S_1$. In the leading order in $\xi$ we find
\bea
S_0=\int_{x_h}^x \frac{x^{\frac{n}{2}}dx}{2x\left[x^n-M x+Ma^2\right]^{\frac{1}{2}}}=
\int_{x_h}^x \frac{\sqrt{{x_h-a^2}}x^{\frac{n}{2}}dx}{2x\left[(x_h-a^2)x^n-(x-a^2)(x_h)^n\right]^{\frac{1}{2}}}
\eea
Here $x_h$ is the value of variable $x$ at the horizon, the point where the denominator vanishes. For $S_1$ we find
\bea
e^{S_1}=\left[\frac{x^{2-n}}{x^n-M x+Ma^2}\right]^{\frac{1}{4}}
\eea
As usual, the WKB approximation breaks down at the ``turning point'', which in our case is located at the horizon, but near that point 
equation  (\ref{PhiXeqnA}) can be approximated as 
\bea
\frac{4x_h[(n-1) x_h-n a^2]}{x_h-a^2}\frac{d}{dx}\left[(x-x_h)\frac{d\Phi}{dx}\right]-\ell(2n+\ell-2)\Phi=0.
\eea
and the solution regular at the horizon is given in terms of the Bessel function $I_0$:
\bea
\Phi=I_0\left[\mu(x-x_h)^\frac{1}{2}\right],\quad \mu=\left\{\frac{\ell(2n+\ell-2)(x_h-a^2)}{x_h[(n-1) x_h-n a^2]}\right\}^\frac{1}{2}\,.
\eea
At large values of $(x-x_h)$ Bessel function behaves as 
\bea\label{WkbBeslAsmp}
I_0\left[\mu(x-x_h)^\frac{1}{2}\right]\simeq \frac{e^{\mu \sqrt{x-x_h}}}{\sqrt{2\pi\mu} (x-x_h)^\frac{1}{4}}
\left[1+\frac{1}{8\mu(x-x_h)^\frac{1}{2}}+\dots\right]\,.
\eea
This matches (\ref{PhiWKB}) with $c_2=0$. To capture the $c_2$ term, which is highly subleading in the Bessel function and in the WKB approximation, one has to introduce a third region which overlaps with both the WKB and Bessel solutions. Rather than doing this, we will use a shortcut by utilizing an analytic continuation. Regularizing the asymptotic expansion for the Bessel function by stepping off the real line and performing the Borel summation, one finds
\bea
I_0(z\pm i0)\simeq \frac{e^z}{\sqrt{2\pi z}}\sum_{k=0}^\infty \frac{[(2k-1)!!]^2}{k!(8z)^k}+
\frac{e^{-z\pm i\pi/2}}{\sqrt{2\pi z}}\sum_{k=0}^\infty \frac{[(2k-1)!!]^2}{k!(-8z)^k}\,.
\eea
This implies the relations $c_2=e^{\pm i\pi/2}c_1$ between the parameters in (\ref{PhiWKB}), as long as one stays off the real axis for $(x-x_h)^\frac{1}{2}$. Therefore, on that axis constants $c_1$ and $c_2$ have the same order of magnitude, i.e., $c_2=\sigma c_1$, where $\sigma$ settles to an $\ell$--independent constant for large values of $\ell$. This leads to our final expression for the WKB approximation (\ref{PhiWKB}):
\bea\label{PhiWKBfin}\label{eqn49}
\Phi\simeq c_1\left[\frac{x^{2-n}}{x^n-M x+Ma^2}\right]^{\frac{1}{4}}\left[e^{(\ell+n-1) S_0}+\sigma e^{-(\ell+n-1) S_0}\right]
\eea
To extract the Love numbers, we need to analyze $S_0$ for large values of $x$. Writing this function as 
\bea\label{eqn410}
S_0(x)=\frac{1}{2}\ln\frac{x}{x_h}+\int_{x_h}^x \left[\frac{\sqrt{{x_h-a^2}}x^{\frac{n}{2}}}{2x\left[(x_h-a^2)x^n-(x-a^2)(x_h)^n\right]^{\frac{1}{2}}}-\frac{1}{2x}\right]dx,
\eea
we observe that the integral converges if the limit is sent to infinity, thus the leading contributions to the wavefunction at large 
values of $x$ can be written as
\bea\label{PhiWKBfinA}
\Phi&\simeq& c_1\left[\left(\frac{x}{x_h}\right)^{\ell/2}e^{\bar S}+\sigma 
\left(\frac{x}{x_h}\right)^{1-n-\ell/2}e^{-{\bar S}}\right]\\
{\bar S}_\ell&=&\frac{\ell+n-1}{2}\int_{1}^\infty \left[\frac{\sqrt{{1-\beta}}z^{\frac{n}{2}}}{z\left[(1-\beta^2)z^n-(z-\beta^2)\right]^{\frac{1}{2}}}-\frac{1}{z}\right]dz\nonumber
\eea
Here we defined a dimensionless quantity $\beta=a^2/x_h$ characterizing the rotation of a black hole. Therefore, the Love number in the WKB approximation is given by
\bea\label{WkbLove}
\la_\ell=\sigma (x_h)^{n+\ell-1}e^{-2{\bar S}_\ell}
\eea
Equations (\ref{PhiWKBfinA}) and (\ref{WkbLove}) constitute our final result for the Love numbers in the WKB approximation. 

\bigskip

While the integral appearing in (\ref{PhiWKBfinA}) cannot be evaluated analytically for arbitrary values of $(n,\beta)$, it is instructive to compute it in the static case and compare the answer with the results obtained in section \ref{SecStatic}. Setting $\beta$ to zero, we find
\bea
{\bar S}_\ell&=&\frac{\ell+n-1}{2}\int_{1}^\infty \left[\frac{z^{\frac{n}{2}}}{z\left[z^n-z\right]^{\frac{1}{2}}}-\frac{1}{z}\right]dz
=\frac{\ell+n-1}{n-1}\left[i\arcsin(z^{(n-1)/2})-\frac{n-1}{2}\ln z\right]_1^\infty\nn
&=&\frac{\ell+n-1}{n-1}\left[\ln\left\{\sqrt{z^{1-n}-1}+i\right\}\right]_1^\infty=\frac{\ell+n-1}{n-1}\ln 2
\nonumber
\eea
This implies that at large values of $\ell$ the Love numbers scale as 
\bea\label{WBBstat1}
\la_\ell\propto M^{\frac{n+\ell-1}{2(n-1)}}2^{-\frac{2\ell}{n-1}}
\eea
Let us compare this with the exact expression (\ref{LoveStaticAnsw}):
\bea\label{WBBstat2}
\la_\ell=
\frac{(k+\frac{1}{2})\tan{\pi k}}{\pi}\frac{\Gamma[1+k]^4}{\Gamma[2(1+k)]^2}M^{2k+1}\propto M^{2k+1}2^{-4k}
\eea
We have used the Stirling approximation and assumed that $\tan{\pi k}$ is of order one for generic values of $k$. Recalling that 
\bea
k=\frac{\ell}{d-3}=\frac{\ell}{2(n-1)}\,,
\eea
we find the perfect match between (\ref{WBBstat1}) and (\ref{WBBstat2}). A similar agreement between (\ref{PhiWKBfinA}) and the exact solution in five dimensions is demonstrated in the Appendix \ref{AppWBK}. Unfortunately, the explicit form of the integral (\ref{PhiWKBfinA}) in the presence of the rotation parameter is not very illuminating.

\section{Perturbations with angular dependence}
\label{SecAngles}

In previous sections we have focused on the Love numbers for the modes with trivial dependence on cyclic coordinates. Let us now discuss the new features of the modes with non--zero quantum numbers corresponding to the cyclic angular directions. After introducing the relevant equations, we will constructing explicit solutions in five dimensions in section \ref{SecN1d5} before making general comments about higher dimensions in section \ref{SecN1gen}.

\bigskip

Let us recall the system of differential equations (\ref{MPodeOdd}) describing separable modes of a massless scalar field in odd dimensions:
\bea\label{CyclEqnOne}
&&r\frac{d}{dr}\left[\frac{\Delta}{r}\frac{d\Phi}{dr}\right]+\frac{R^2}{\Delta}
\left[
\sum_k\frac{a_k n_k}{r^2+a_k^2}\right]^2\Phi-\Omega\Phi-r^2 P_{n-2}[r^2]\Phi=0,\nn
&&{x_i}\frac{d}{dx_i}\left[\frac{H_i}{x_i}\frac{dX_i}{dx_i}\right]-
{H_i}{}\left[
\sum_k\frac{a_k n_k}{a_k^2-x_i^2}\right]^2X_i+\Omega X_i-x_i^2P_{n-2}[-x_i^2]X_i=0\,, \\
&&\Omega=\left[\prod a_i\right]^2\left[\sum_k\frac{n_k}{a_k}\right]^2\,.\nonumber
\eea
We removed time dependence but kept quantum numbers associated with angular directions. To take the limit of equal rotations, we have to repeat the arguments presented in section \ref{SecEqnSetup}. Assuming that the rotation parameters and angular coordinates are ordered as (\ref{RangeMPodd}), introducing expansions (\ref{RangeMPexp1})--(\ref{RangeMPexp2}) with $k=n$, 
and sending $\eps$ to zero while keeping $(b_s,y_s)$ fixed, one can expand the angular equation for $X_1(y_1)$ in powers of $\eps$. 
Then the first fer powers in the terms with $X_1(y_1)$ cannot be matched by contributions from derivatives, and this leads to constraints of the polynomial $P_{n-2}[-x_i^2]$. Specifically, one finds
\bea\label{PolynPn1}
\Omega-x^2P_{n-2}[-x^2]=(a^2-x^2)^{n-1}p^2-\Lambda x^2 a(a^2-x^2)^{n-2},\quad p=\sum n_k.
\eea
This relation generalizes the relation (\ref{PolynMPaEq}) to the situation with nontrivial dependence on cyclic coordinates. In the resulting differential equation for $X_n(y_n)$, constant $a$ is the only dimensionfull parameter, so the separation constant 
$\Lambda$ cannot depend on it. Thus the eigenvalue $\Lambda$ comes from the standard spherical harmonics in $d$ dimensions, and it is given by
\bea
\Lambda=\ell (\ell+d-3).
\eea
Substituting the polynomial $P_{n-2}$ defined by (\ref{PolynPn1}) into the radial portion (\ref{CyclEqnOne}) with equal rotations, 
we arrive at the final radial equation:
\bea\label{CyclEqnTwo}
r\frac{d}{dr}\left[\frac{\Delta}{r}\frac{d\Phi}{dr}\right]+\frac{\rho^{4n-4}}{\Delta}
(ap)^2\Phi-p^2\rho^{2n-2}\Phi-\Lambda r^2\rho^{2n-4}\Phi=0,\quad \rho^2=r^2+a^2.
\eea
This equation will be analyzed in section \ref{SecN1gen}, but before doing that we will discuss the radial portion of (\ref{CyclEqnOne}) with arbitrary rotation parameters $(a_1,a_2)$ in five dimensions.

\subsection{Explicit solutions in five dimensions}
\label{SecN1d5}

In this subsection we will analyze separable solutions of the wave equation in five dimensions. This problem was studied in \cite{RSST}, where the authors wrote the solution in terms of hypergeometric functions and extracted the corresponding Love numbers. This construction is reviewed in Appendix \ref{App5d}. Unfortunately, the solution found in \cite{RSST} does not reduce to the general form (\ref{PhiHyper}) of solutions for static geometries in arbitrary dimensions, so we introduce a different parameterizarion that cures this problem. Furthermore, the construction introduced in \cite{RSST} fails for extremal black holes, and we fill this gap as well.

\bigskip

Let us consider the radial equation from the system (\ref{CyclEqnOne}) in five dimensions. In this case, the polynomial $P_{n-2}$ is just a constant, and the equation for function $\Phi(r)$ can be rewritten as 
\bea\label{eqn5aOneMn}
r\frac{d}{dr}\left[\frac{\Delta}{r}\frac{d\Phi}{dr}\right]+\frac{F r^2}{\Delta} \Phi-\Lambda r^2\Phi=0,
\eea
where function $F$ simplifies to
\bea\label{eqn5aTwoMn}
F=M (bn_\phi+an_\psi)^2 +(a^2-b^2)\Big[(n_\phi)^2 R_b-(n_\psi)^2 R_a\Big].
\eea
and the separation constant is parameterized as
\bea\label{5dLamMn}
\Lambda=\ell(\ell+2)\quad\mbox{with integer}\quad \ell.
\eea 
To avoid unnecessary clutter from double subscripts, we modified the notation for angles and rotation parameters as 
\bea
\phi_1=\phi,\quad \phi_2=\psi,\quad a_1=a,\quad a_2=b.
\eea

\bigskip
\noindent
{\bf Review of a known construction}

The solution of equation (\ref{eqn5aOneMn}) which remains regular at the horizon was analyzed in \cite{RSST}, and this construction is reviewed in the Appendix \ref{App5dKnown}. The result reads (\ref{eqn5dSoln})
\bea\label{eqn5dSoln}
\Phi=z^{\ell/2+1}(1-z)^{i({m_L}+{m_R})}F\Big[\mathfrak{a},\mathfrak{b},1+\mathfrak{a}+\mathfrak{b}-\mathfrak{c},1-z\Big].
\eea
Various constants entering this expression are given by
\bea
&&\mathfrak{a}=1+\frac{\ell}{2}+2im_L,\quad \mathfrak{b}=1+\frac{\ell}{2}+2im_R,\quad \mathfrak{c}=1+{\ell}\,.\\
&&{m_L}=\frac{a-b}{r_+  + r_-}\frac{n_\phi-n_\psi}{4} \quad {m_R}=\frac{a+b}{r_+  - r_-}\frac{n_\phi+n_\psi}{4}\,,
\eea
where $r_+$ and  $r_-$ are the radii of the outer and inner horizons. 
The variable $z$ appearing in (\ref{eqn5dSoln}) was defined by the authors of \cite{RSST}  through two changes of coordinates:
\bea\label{RsqRodrMn}
r^2 = \frac{(r_+^2-r_-^2)x + r_{+}^2+r_{-}^2}{2}\,,\quad z=\frac{2}{1+x}\,.
\eea
As expected from the general analysis presented in section \ref{SecPertSpec}, at large values of $r$, the solution (\ref{eqn5dSoln}) behaves as 
(\ref{PsiAsympGen}) with $d=5$:
\bea\label{PsiAsympD5}
\Phi=b_1 r^\ell\Bigg\{\sum_{s=0}^{k+1}\frac{M^s}{r^{2s}} f_s(r)+
r^{-2-2\ell}\left[b_2\phi_2+
b_3 \phi_3 \ln r^2\right] \Bigg\}\,.
\eea
with $f_0(0)=1$. The values of $(b_2,b_3)$ follow from the expansion of (\ref{eqn5dSoln}) near $z=0$, and they can be found in \cite{RSST}. For example, the parameter specifying the running is given in terms of the Pochhammer symbols 
\bea
b_3= -(-1)^{l}\frac{(-l + 2i{m_{L}})_{l+1}(-l +2 i {m_{R}})_{l+1}}{l!(l+1)!}\,.
\eea
The expression for $b_2$ is more complicated.

\bigskip
\noindent
{\bf New parameterization of the solution}

It is useful to find an alternative form of the solution (\ref{eqn5dSoln}) which reduces to the general expression (\ref{PhiHyper}) in the static case. The relevant derivation is performed in the Appendix \ref{App5dNew}, and here we just summarize the results.

Introducing a new coordinate 
\bea\label{Xhyper1Mn}
x=a_1 r^2+a_2\,,
\eea
writing the wavefunction as 
\bea
\Phi=x^{p_1}(x-1)^{p_2}g\left[1-\frac{1}{x}\right].
\eea
and requiring (\ref{eqn5aOneMn}) to reduce to the hypergeometric equation for $g[z]$, we find the values of the parameters 
$(a_1,a_2,p_1,p_2)$:
\bea\label{Xhyper2Mn}
&&a_2 = \frac{1}{2} \left[1 + ( a^2 + b^2-M) a_1\right], \quad
a_1 = \frac{1}{\sqrt{(M-a^2-b^2)^2-4(a b)^2}}\,,\nn
&&p_1=\frac{\ell}{2}-p_2,\quad p_2=\frac{1}{4}\left[\frac{(a+b)(n_\phi+n_\psi)}{\sqrt{M-(a+b)^2}}+\frac{(a-b)(n_\phi-n_\psi)}{\sqrt{M-(a-b)^2}}\right]\,.
\eea
Then the solution regular at at the horizon reads (see (\ref{5dHyperSoln})):
\bea\label{5dHyperSolnMn}
&&\Phi=x^{\ell/2}\left[1-\frac{1}{x}\right]^{p_2}F\left[A-B,A+B,1+2p_2,1-\frac{1}{x}\right],\nn 
&&A=p_2-\frac{\ell}{2},\quad B=\left[(p_2)^2+\frac{a_1}{4}(a^2-b^2)(p^2-q^2)\right]^{\frac{1}{2}}\,,
\eea
If the angular quantum numbers vanish ($n_\phi=n_\psi=0$), then this expression simplies to 
the relation
\bea
\Phi=x^{\ell/2}F\left[-\frac{\ell}{2},-\frac{\ell}{2},1,1-\frac{1}{x}\right]\,,
\eea
which in turn reduces to (\ref{PhiHyper}) with $d=5$ when rotation parameters are set to zero. 

\bigskip
\noindent
{\bf Extremal black holes}

The analysis presented above works for non--extremal black holes, but it breaks down in the extremal limit where\footnote{To avoid unnecessary complications, we assume that both rotation parameters are non--negative. The general extremal limit is $M=(|a|+|b|)^2$.} $M=(a+b)^2$. In this case, the changes of variables (\ref{RsqRodrMn}) and (\ref{Xhyper1Mn}), (\ref{Xhyper2Mn}) become singular. This problem is obvious in (\ref{Xhyper2Mn}), and for (\ref{RsqRodrMn}) one has to recall that two horizons located at $r_+$ and $r_-$ coincide in the extremal case where $\Delta=(r^2-ab)^2$. The extremal limit of equations (\ref{eqn5aOneMn})--(\ref{eqn5aTwoMn}) is analyzed in the Appendix \ref{App5dExtrem}, and here we just summarize the results.

Writing the wavefunction $\Phi$ as 
\bea
\Phi(r)=\sqrt{z}Z(z),\quad z=\frac{b_1}{r^2-ab}\,,
\eea
where $b_1$ is an arbitrary length scale, and solving the resulting equation for $Z(z)$, one 
finds the expression in terms of the Kummer's confluent hypergeometric function
\bea\label{ExtrCHFmain}
Z&=&e^{-i\Xi\, z/2}\left\{c_1 z^{\frac{1+\ell}{2}}
\mathfrak{M}\left[iA+1+\frac{\ell}{2},2+\ell,i\Xi z\right]+
c_2 z^{-\frac{1+\ell}{2}}\mathfrak{M}\left[iA-\frac{\ell}{2},-\ell,i\Xi z\right]
\right\},\nn
A&=&\frac{(a^2-b^2)(n_\phi^2-n_\psi^2)}{4b_1\Xi},\quad \Xi=\frac{a+b}{b_1}\left[(a^2+b^2)(n_\phi^2+n_\psi^2)-ab(n_\phi-n_\psi)^2\right]
\eea
The solution regular at the horizon is given it terms of the Tricomi function $U$,
\bea\label{TricomiMn}
Z=e^{-i\Xi\, z/2}z^{\frac{1+\ell}{2}}U\left[iA+1+\frac{\ell}{2},2+\ell,i\Xi z\right]
\eea
and as reviewed in the Appendix \ref{App5dExtrem}, this implies the relation between coefficients $(c_1,c_2)$ in (\ref{ExtrCHFmain}):
\bea
\frac{c_2}{c_1}=\frac{\Gamma(\bb-1)}{\Gamma(\aa)}\, 
\frac{\Gamma(\aa-\bb+1)}{\Gamma(1-\bb)},\quad \aa=iA+1+\frac{\ell}{2},\quad \bb=2+\ell\,.
\eea
This ratio determines the Love number. The wavefunction $Z$ simplifies 
for $a=b$ or for $n_\psi=\pm n_\psi$, and the result can me written in terms of the Bessel functions
\bea
Z=J_{(\ell+1)/2}\left[\frac{{\Xi}\,z}{2}\right]-i Y_{(\ell+1)/2}\left[\frac{{\Xi}\,z}{2}\right].
\eea
We refer to the Appendix \ref{App5dExtrem} for details. 

\subsection{Comments on excitations in higher dimensions}
\label{SecN1gen}

Although we will not attempt to compute Love numbers for excitations with arbitrary angular quantum numbers in higher dimensions, it is instructive to discuss the boundary conditions for such modes. Since extremal and non--extremal black holes lead to drastically different behaviors of the scalar at the horizon, we will discuss these two cases separately. As reviewed in the Appendix \ref{App4d}, in four dimensions, the boundary conditions are imposed by the physical requirement for having only the ingoing waves at the horizon. A similar picture would work in higher dimensions as well, but interestingly, already the mathematical requirement of normalizability implies that one can have either pure ingoing or purely outgoing waves, while a mixture is not allowed. We will begin with showing this for non--extremal black holes, and conclude with summarizing the results for a more involved extremal case, which is presented in the Appendix \ref{AppExtrBC}.

\bigskip
\noindent
Let us analyze equation (\ref{CyclEqnTwo}) in the vicinity of the horizon assuming that the black hole is not extremal. We begin with rewriting this this equation in terms of $\rho=a^2+r^2$,
\bea\label{CyclEqnTwoX}
4\frac{d}{d\rho}\left[{\Delta}\frac{d\Phi}{d\rho}\right]+\frac{\rho^{2n-2}-a^{2n-4}\Delta}{\Delta(\rho-a^2)}
(ap)^2\Phi-\Lambda \rho^{n-2}\Phi=0.
\eea
We recall that $\Delta=\rho^n-M(\rho-a^2)$. 
Equation (\ref{CyclEqnTwoX}) can be obtained from an effective action,
\bea\label{CyclEqnTwoXact}
S=\int d\rho\left[4{\Delta}\left|\frac{d\Phi}{d\rho}\right|^2-\frac{\rho^{2n-2}-a^{2n-4}\Delta}{\Delta(\rho-a^2)}
(ap)^2|\Phi|^2+\Lambda \rho^{n-2}|\Phi|^2\right],
\eea
and it is natural to focus only on solutions for which this action is normalizable. Specifically, we will analyze normalizability in the vicinity of the horizon. 

The horizon, $\rho=\rho_h$, is located at the points where $\Delta=0$, and this leads to a complicated equation for $\rho_h$. One can avoid this problem by solving the equation 
\bea
\Delta(\rho_h)=\rho_h^n-M(\rho_h-a^2)=0\nonumber
\eea
for $M$  and using $\rho_h$ as a parameter controlling the mass of the solution. This gives the expression for $\Delta$ which will be used below:
\bea
\Delta=\rho^n-\frac{\rho-a^2}{\rho_h-a^2}\rho_h^n\,.
\eea
The general solution of equation  (\ref{CyclEqnTwoX}) has the form
\bea\label{CyclEqnTwoXsoln}
\Phi=c_1(\rho - \rho_h)^{i\beta} \left[1+\sum d_k (\rho - \rho_h)^k\right]+
c_2(\rho - \rho_h)^{i\beta} \left[1+\sum {\tilde d}_k (\rho - \rho_h)^k\right]
\eea
where $\beta$ is defined by
\bea
\beta=\frac{a p \sqrt{\rho_h-a^2}}{2 [n \rho_h-\rho_h-a^2 n]}
\eea
Substitution of the solution (\ref{CyclEqnTwoXsoln}) into the action (\ref{CyclEqnTwoXact}), 
one concludes that normalizability requires the product $(c_1c_2)$ to vanish, so the solution reads
\bea\label{CyclEqnTwoXsoln2}
\Phi=c_1(\rho - \rho_h)^{i\beta} \left[1+\sum d_k (\rho - \rho_h)^k\right].
\eea
Alternatively, one could take the complex conjugate of (\ref{CyclEqnTwoXsoln2}), but a discrete choice has to be made. Physically (\ref{CyclEqnTwoXsoln2}) corresponds to an ingoing wave, and the complex conjugate corresponds to an outgoing one. 

\bigskip

The analysis of the extremal case is more involved since the integrand of (\ref{CyclEqnTwoXact}) has a higher order pole, so several terms in the Laurent have to cancel, and the relevant analysis is performed in the Appendix \ref{AppExtrBC}. At the leading order in the vicinity of the horizon, the solution is given by (\ref{PhiNdepOne}), 
\bea\label{PhiNdepOneMn}
{\Phi}\simeq c_1 \exp\left[\frac{ia p}{2 (n-1) {z}}\right] +c_2
\exp\left[-\frac{ia p}{2 (n-1) {z}}\right],\quad z=r-r_h
\eea
In the Appendix \ref{AppExtrBC}, we construct several subleading terms as well and demonstrate that 
the corrected function $\Phi$ becomes normalizable in (\ref{CyclEqnTwoXact}) if and only if the product 
$(c_1c_2)$ vanishes. This again selects ingoing or outgoing boundary conditions. For such normalizable solutions one finds
\bea\label{NormAnlzPhiMn}
{\Phi}=c_1 \exp\left[\frac{ia p}{2 (n-1) {z}}-\frac{(n-1) p}{6 \sqrt{n_-} n}\ln{z}\right]\left(1 +
   u_1 z+u_2 z^2+\dots\right),
\eea
where $(u_1,u_2)$ are complicated numerical coefficients computed in the Appendix \ref{AppExtrBC}.

\section{Discussion}

Let us begin with summarizing the results of this article. We clarified the relationship between the Love numbers and logarithmic running, and we demonstrated that the latter is not sensitive to the behavior of the perturbation at the horizon, as it takes the same values for regular and singular solutions. This implies that even in the presence of logarithmic running, the informative Love numbers are given by the coefficients in front of pure powers, in contrast to some statements made in the literature. We found the exact formulas for the logarithmic runnings in all dimensions, as well as approximate expressions for the genuine Love numbers in several distinct regimes.

\bigskip

Previous calculations of Love numbers have relied on explicit solutions of equations of motion which remained regular at the horizon, and they distinguished the cases of specific numbers and logarithmic running. We have demonstrated that the running is not sensitive to the boundary conditions at the horizon, and thus it knows about black holes only through the asymptotic behavior of the metric at infinity. It is the constant contribution that maintains some freedom before the boundary conditions at the horizon are imposed, and thus it is that constant rather than the running coefficient that should be interpreted as the proper Love number. On the other hand, evaluation of logarithmic running is easier since one does not have to look at the entire geometry, and in section \ref{SecLogRunSec} we computed such running coefficients for Myers--Perry black holes in all dimensions. Evaluation of the genuine Love numbers, which are determined by fixing integration constants in the solutions, requires the knowledge of the wavefunction everywhere, and thus we carried it out only for special cases and in some approximations. The results for non--extremal black holes in five dimensions were known, although in section \ref{SecAngles} we cast them in a slightly different form (\ref{5dHyperSolnMn}), which in the static case naturally reduces to the solutions (\ref{PhiHyper}) for the Tangherlini geometry applicable to all dimensions. In the extremal case, we constructed the new exact solutions for perturbations and used them to extract the Love numbers in five and seven dimensions. This was done in sections \ref{SecAngles} and \ref{SecExtremal}. 

While the wave equation can be solved exactly in the static case and for rotating black holes in four, five and seven dimensions, unfortunately analytical solutions for the general Myers--Perry geometry are beyond reach.  To circumvent this problem, in section \ref{SecSlow} we calculated Love numbers using perturbation theory in the rotation parameter. We found that, with few exceptions, even Love numbers vanishing for static  black holes acquire a nontrivial dependence on the rotation parameter once it is introduced. Therefore, vanishing of the Love numbers for rotating geometries appears to be a peculiarity of the four--dimensional case.

In section \ref{SecWKB} we applied a different approximation and computed the leading asymptotic behavior of Love numbers for large angular momentum of the excitation, while keeping the rotation parameters of black holes unrestricted. We also confirmed the accuracy of this WKB approximation by comparing it with the exact answers for static black holes and for the Myers--Perry geometry in five dimensions.

\bigskip

Our results can be extended in several directions. First, while we focused on scalars, one may study equations for particle with spin, and some results for rotating black holes in specific dimensions have already appeared in \cite{LoveKerrM3,StatLNgraviton,LoveKerrM4f}. Second, one can consider perturbations of charged black holes rather than neutral Myers--Perry geometries, and some results for such excitations have been obtained in \cite{LoveCharge4,LoveCharge1,LoveCharge2,LoveCharge3}. Finally, it would be very interesting to connect our results to the description of Love numbers in terms of effectine field theories advocated in \cite{StaticLN1,StaticLN2,LoveKerrM3,LoveRotHd1,LNeft0,LNeft1,LNeft2,LNeft3}.

\appendix

\section{Myers--Perry black holes and their excitations}
\label{AppMP}

In this appendix we will summarize some well-known properties of the Myers--Perry geometry \cite{MyersPerry}. Specifically, we will present the metrics, the Killing--Yano tensors, the special frames leading to separation of variables in various dynamical equations, and the details this separation for the wave equation. Since the black hole geometries have different structures in even and odd dimensions, these two cases will be discussed separately.

\bigskip

\noindent
{\bf Even dimensions}

We begin with quoting the Myers--Perry solution in $d=2n+2$ dimensions \cite{MyersPerry,Myers}:
\bea\label{MPeven}
ds^2&=&-dt^2+\frac{Mr}{FR}\Big(dt+\sum_{i=1}^n a_i\mu_i^2 d\phi_i\Big)^2+\frac{FR dr^2}{R-Mr}
+\sum_{i=1}^n(r^2+a_i^2)\Big(d\mu_i^2+\mu_i^2 d\phi_i^2\Big)\nn
&&+r^2d\alpha^2.
\eea
Here variables $(\mu_i,\alpha)$ are subject to a constraint
\bea
\alpha^2+\sum_{i=1}^n\mu_i^2=1,
\eea
and functions $F$, $R$ are defined by 
\bea\label{MPdefFR}
F=1-\sum_{k=1}^n\frac{a_k^2\mu_k^2}{r^2+a_k^2},\quad R=\prod_{k=1}^n (r^2+a_k^2).
\eea
The geometry (\ref{MPeven}) admits $(n+1)$ Killing vectors corresponding to translations in $(t,\phi_i)$ directions, as well as a family of anti--symmetric Killing--Yano 
tensors (KYT) \cite{Kub1w1,Kub1w2,Kub1w3,Kub1w4,Kub1w5,Kub1w6,Kub2w1,Kub2w2,
Kub2w3,Kub3w1,
Kub3w2,Kub3w3,ChLKill,KubRev}, which satisfy differential equations \cite{YanoW1,YanoW2,YanoW3}
\bea
\nabla_\mu Y_{\nu_1\dots \nu_p}+\nabla_{\nu_1} Y_{\mu\nu_2\dots \nu_p}=0,
\eea
To describe such tensors, it is convenient to replace $(r,\mu_1\dots \mu_n)$ by new coordinates $(x_0,x_1\dots x_n)$ defined by
\bea\label{Xmap}
x_0=r^2,\quad (a_i\mu_i)^2=\frac{1}{c_i^2}\prod_1^n(a_i^2-x_k^2),\quad c_i^2=\prod(a_i^2-a_k^2).
\eea
The rotation parameters and the ellipsoidal coordinates $x_j$ can be ordered as 
\bea\label{XAorderEven}
x_n\le a_n\le \dots x_1 \le a_1\,.
\eea
All Killing--Yano tensors corresponding to the Myers--Perry black holes (\ref{MPeven}) can be summarized by a very elegant formula \cite{Kub1w1,Kub1w2,Kub1w3,Kub1w4,Kub1w5,Kub1w6,Kub2w1,Kub2w2,Kub2w3,Kub3w1,
Kub3w2,Kub3w3}
\bea\label{Kub2}
Y^{2(n-k)}=\star\left[\wedge h^k\right],
\eea
where $h$ is a two--form that can be written as 
\bea\label{HInFrame}
h=re^r\wedge e^t+\sum_i x_i e^{x_i}\wedge e^i\,.
\eea
Here $e^A$ are special frames,
\bea
ds^2=-(e^t)^2+(e^r)^2+\sum_k [(e^{x_k})^2+(e^k)^2],
\eea
constructed in \cite{Kub1w1,Kub1w2,Kub1w3,Kub1w4,Kub1w5,Kub1w6,Kub2w1,Kub2w2,Kub2w3,Kub3w1,
Kub3w2,Kub3w3,UniqueKYw1,UniqueKYw2,
UniqueKYw3,UniqueKYw4}, and we will write only the components with upper indices, $e_A^\mu$, using the notation introduced in \cite{ChLKill}:
\bea\label{AllFramesMP}
e_t&=&-\sqrt{\frac{R^2}{FR\Delta}}\left[\d_t-
\sum_k\frac{a_k}{r^2+a_k^2}\d_{\phi_k}\right],\quad 
e_r=\sqrt{\frac{\Delta}{FR}}\d_r,\nn
e_i&=&-\sqrt{\frac{H_i}{d_i}}\left[\d_t-\sum_k\frac{a_k}{a_k^2-x_i^2}\d_{\phi_k}
\right],\quad e_{x_i}=\sqrt{\frac{H_i}{d_i}}\d_{x_i}\,.
\eea
Here we introduced convenient expressions involving the $x$ coordinates defined by (\ref{Xmap}) and the radial parameter $r$,
\bea\label{MiscElliptic}
&&H_i=\prod_k^n(a_k^2-x_i^2),\quad d_i=(r^2+x_i^2)\prod_{k\ne i}^n(x_k^2-x_i^2)\,,\nn
&&R=\prod_{k}^n (r^2+a_k^2),\quad
FR=\prod_k^n(r^2+x_k^2),\quad \Delta=R-Mr.
\eea
Functions $(R,FR)$ can be interpreted as $H_0$ and $d_0$ with $x_0=i r$. Note that, apart from the common overall factors, the components $(e_t^\mu,e_r^\mu)$ of the frames depend only on $r$, while the components $(e_i^\mu,e_{x_i}^\mu)$ depend only on $x_i$. This crucial fact is responsible for separation of variables in dynamical equations for scalars \cite{Kub1w1,Kub1w2,Kub1w3,Kub1w4,Kub1w5,Kub1w6,Kub2w1,Kub2w2,Kub2w3,Kub3w1,
Kub3w2,Kub3w3,UniqueKYw1,UniqueKYw2,
UniqueKYw3,UniqueKYw4}, vectors \cite{LMaxw,KubMaxwW1,KubMaxwW2,KubMaxw1w1,KubMaxw1w2,KubMaxw1w3}, higher antisymmetric forms \cite{LBfield}, and gravitons \cite{LGravWave}. We will focus on the equation for the massless scalar.

\bigskip
\noindent
Let us consider the wave equation,
\bea\label{WaveEqn}
\frac{1}{\sqrt{-g}}\d_\mu\left[\sqrt{-g}g^{\mu\nu}\d_\nu \Psi\right]=0,
\eea
in the Myers--Perry background (\ref{MPeven}). Imposing a separable ansatz
\bea\label{WaveEqnTwo}
\Psi=E\Phi(r)\left[\prod X_i(x_i)\right],\qquad E=e^{i\omega t+i\sum n_i\phi_i}\,,
\eea
and substituting in into (\ref{WaveEqn}), we conclude that the function $\Psi$ satisfies a system of ordinary differential equations \cite{LMaxw}:
\bea\label{SeparWEKerrA}
\label{MPodeEven}
&&\frac{\d}{\d r}\left[\Delta \frac{\d\Psi}{\d r}\right]+\frac{R^2}{\Delta}
\left[\omega-
\sum_k\frac{a_k n_k}{r^2+a_k^2}\right]^2\Psi-P_{n-1}[-r^2]\Psi=0\,,\nn
&&\frac{\d}{\d x_i}\left[H_i \frac{\d \Psi}{\d x_i}\right]-
{H_i}\left[\omega-
\sum_k\frac{a_k n_k}{a_k^2-x_i^2}\right]^2\Psi+P_{n-1}[x_i^2]\Psi=0\,,\\
&&\d_t\Psi=i\omega\Psi,\quad \d_{\phi_k}\Psi=in_k\Psi\,.\nonumber
\eea
Here $P_{n-1}[y]$ is an arbitrary polynomial of degree $(n-1)$, and it is crucial that all equations (\ref{SeparWEKerrA}) contain {\it the same} function $P_{n-1}$. Therefore, the separable solutions of equation (\ref{WaveEqn}) satisfy a system of $d=2(n+1)$ ordinary differential equations for function $\Psi$, and as expected, the result depends on $d-1=2n+1$ integration constants:
\label{CountWave}
\begin{itemize}
\item $(n+1)$ parameters $(\omega,n_i)$;
\item $n$ free coefficients of the polynomial $P_{n-1}$.
\end{itemize}
This guarantees full integrability of the wave equation. To prove integrabilty of equations for vector and tensor fields, one has to identify the projections that become separable, and in the case of the vector field, the most general separable solutions were found in \cite{LMaxw}, where it was shown that the projections of the vector onto the special frames (\ref{AllFramesMP}) must have the form\footnote{See also \cite{KubMaxwW1,KubMaxwW2,KubMaxw1w1,KubMaxw1w2,KubMaxw1w3} for further explorations of this ansatz.}
\bea\label{MaxwAnstz}
[m^{(I)}_\pm]^\nu A_\nu=\left[\zeta+\frac{x_I}{x_I\pm\mu}\right][m^{(I)}_\pm]^\nu\d_\nu \Psi,\quad x_0=-ir,
\eea
where
\bea
m_\pm^{(0)}=\sqrt{FR}(e_r\mp e_t),\quad m_\pm^{(j)}=\sqrt{d_j}(e_{x_j}\mp i e_j).
\eea
Function $\Psi$ satisfies a modified version of the system (\ref{SeparWEKerrA}), and we refer to 
\cite{LMaxw} for details. Similar projections onto the frames (\ref{AllFramesMP}) lead to separability of equations for $p$--form fields \cite{LBfield} and for gravitons \cite{LGravWave}.

\bigskip

\noindent
{\bf Odd dimensions}

Let us now turn to the Myers--Perry black holes in odd dimensions ($d=2n+1$). The metric is given by \cite{MyersPerry,Myers}
\bea\label{MPodd}
\hskip -0.3cm
ds^2=-dt^2+\frac{Mr^2}{FR}\Big(dt+\sum_{i=1}^{n} a_i\mu_i^2 d\phi_i\Big)^2+\frac{FR dr^2}{R-Mr^2}
+\sum_{i=1}^{n}(r^2+a_i^2)\Big(d\mu_i^2+\mu_i^2 d\phi_i^2\Big),
\eea
and coordinates $\mu_i$ are subject to a constraint
\bea
\sum_{i=1}^{n}\mu_i^2=1.\nonumber
\eea
The counterparts of the special frames (\ref{HInFrame}) are given by \cite{Kub1w1,Kub1w2,Kub1w3,Kub1w4,Kub1w5,Kub1w6,ChLKill}
\bea\label{AllFramesMPOdd}
e_t&=&-\sqrt{\frac{R^2}{FR\Delta}}\left[ \d_t
-\sum_k\frac{a_k}{r^2+a_k^2}\d_{\phi_k}\right],\quad e_r=\sqrt{\frac{\Delta}{FR}}\d_r,\quad
e_{x_i}=\sqrt{\frac{H_i}{x_i^2 d_i}}\d_{x_i}\nn
e_i&=&-\sqrt{\frac{H_i}{x^2_id_i}}\left[\d_t-\sum_k\frac{a_k}{a_k^2-x_i^2}\d_{\phi_k}
\right],\quad 
e_\psi=-\frac{\prod a_i}{r\prod x_k}\left[\d_t-\sum_k\frac{1}{a_k}\d_{\phi_k}
\right].
\eea
Now there is an additional frame $e^\psi$, and the metric has the form
\bea
ds^2=-(e^t)^2+(e^r)^2+\sum_k [(e^{x_k})^2+(e^k)^2]+(e^\psi)^2.
\eea
The relations (\ref{MiscElliptic}) are modified, and functions appearing in 
(\ref{AllFramesMPOdd}) are defined by
\bea\label{MiscEllipticOdd}
&&H_i=\prod_k^n (a_k^2-x_i^2),\quad d_i=(r^2+x_i^2)\prod_{k\ne i}^{n-1}(x_k^2-x_i^2)\,,\nn
&&R=\prod_{k}^n (r^2+a_k^2),\quad
FR=r^2 \prod_k^{n-1}(r^2+x_k^2),\quad \Delta=R-Mr^2.
\eea
The rotation parameters and ellipsoidal coordinates $x_j$ are ordered in a way which is slightly different from (\ref{XAorderEven}):
\bea\label{XAorderOdd}
a_{n}\le x_{n-1}\le \dots \le a_1\,.
\eea
As in the even--dimensional case, the frames (\ref{AllFramesMPOdd})) can be used to demonstrate full separability of equations for scalars \cite{Kub1w1,Kub1w2,Kub1w3,Kub1w4,Kub1w5,Kub1w6,Kub2w1,Kub2w2,Kub2w3,Kub3w1,
Kub3w2,Kub3w3,UniqueKYw1,UniqueKYw2,
UniqueKYw3,UniqueKYw4}, vectors \cite{LMaxw,KubMaxwW1,KubMaxwW2,KubMaxw1w1,KubMaxw1w2,KubMaxw1w3}, higher antisymmetric forms \cite{LBfield} and gravitons \cite{LGravWave}

\bigskip
\noindent
To solve the wave equation (\ref{WaveEqn}), one again imposes a separable ansatz (\ref{WaveEqnTwo}), which leads to the system of ordinary differential equations for function 
$\Psi$ \cite{LMaxw}\footnote{While the explicit system of differential equations (\ref{MPodeOdd}) was first written in \cite{LMaxw}, the separability of the wave equation in coordinates $(r,x_i)$ was demonstrated much earlier in  \cite{Kub1w1,Kub1w2,Kub1w3,Kub1w4,Kub1w5,Kub1w6,Kub2w1,Kub2w2,Kub2w3,Kub3w1,
Kub3w2,Kub3w3,UniqueKYw1,UniqueKYw2,
UniqueKYw3,UniqueKYw4}}.: 
\bea\label{MPodeOdd}
&&r\frac{d}{dr}\left[\frac{\Delta}{r}\frac{d\Phi}{dr}\right]+\frac{R^2}{\Delta}
\left[\omega-
\sum_k\frac{a_k n_k}{r^2+a_k^2}\right]^2\Phi-\Omega^2\Phi-r^2 P_{n-2}[r^2]\Phi=0,\\
&&{x_i}\frac{d}{dx_i}\left[\frac{H_i}{x_i}\frac{dX_i}{dx_i}\right]-
{H_i}{}\left[\omega-
\sum_k\frac{a_k n_k}{a_k^2-x_i^2}\right]^2X_i+\Omega^2X_i-(x_i)^2P_{n-2}[-x_i^2]X_i=0\,, \nn
&&\d_t\Psi=i\omega\Psi,\quad \d_{\phi_k}\Psi=in_k\Psi\,.\nonumber
\eea
Here we defined a convenient combination of constants
\bea
\Omega=\left[{\prod a_i}\right]\left[\omega-\sum_k\frac{n_k}{a_k}
\right]\,.
\eea
The system (\ref{MPodeOdd}) has $d-1=2n$ integration constants:
\label{CountWaveOdd}
\begin{itemize}
\item $(n+1)$ parameters $(\omega,n_i)$;
\item $(n-1)$ free coefficients of the polynomial $P_{n-2}$.
\end{itemize}
Therefore, the wave equation in the Myers--Perry background is fully separable. The solutions of the systems (\ref{MPodeEven}) and (\ref{MPodeOdd}) are explored throughout this article.

\section{Love numbers in four dimensions}
\label{App4d}

Although it is well known that in four domensions the Love number vanish for the various excitations of the Kerr black hole \cite{LoveKerrM1,LoveKerrM2,LoveKerrM3,LoveKerrM4f,LoveKerrM0}, it is instructive to summarize the relevant argument for the scalar field utilizing the notation which is used for higher dimensions throughout this article.

\bigskip

For the Kerr black hole, the radial equation in (\ref{MPodeEven}) becomes
\bea\label{EqnKerrWave}
\frac{\d}{\d r}\left[[(r^2+a^2)-M r] \frac{\d\Phi}{\d r}\right]+
\frac{(a n)^2(r^2+a^2)}{(r^2+a^2)-M r}\Phi
-\ell(\ell+1)\Phi=0\,.
\eea
We have set $\omega$ to zero, used the expression (\ref{MiscElliptic}) for $\Delta$, and observed that for time--independent modes, the eigenvalue of the angular equation is given by 
$\Lambda=\ell(\ell+1)$. In the static case, the solution regular at the horizon is given by (\ref{PhiHyper}),
\bea
\Phi=c_1 r^{\ell}F\left[-{\ell},-{\ell},1,1-\frac{M}{r}\right]\,.
\eea
Turning on the rotation parameter $a$, but keeping $n=0$, one finds a minor modification of this solution:
\bea
\Phi=c_1 x^{\ell}F\left[-{\ell},-{\ell},1,1-\frac{1}{x}\right]\,,\quad
x=\frac{r-M/2}{\sqrt{M^2-4a^2}}+\frac{1}{2}.
\eea
In the presence of angular dependence, i.e., for nontrivial $n$, we look for solution in the form
\bea\label{EqnKerrPhi}
\Phi=x^{p_1}(x-1)^{p_2} g\left[1-\frac{1}{x}\right],
\eea
and try to find the values of $(p_1,p_2)$ which lead to a hypergeometric equation for $g(z)$. Substitution of the ansatz (\ref{EqnKerrPhi}) into (\ref{EqnKerrWave}) leads to a differential equation
\bea
z(1-z)g''+(2p_2+1+2p_1 z-z)g'+\frac{G(z)}{z(1-z)}g=0
\eea
The explicit form of polynomial $G(z)$, which depends on $(p_1,p_2)$, is rather complicated, but it leads to the hypergeometric equation only if two conditions are satisfied:
\bea
G(0)=0,\quad G(1)=0,\nonumber
\eea
This gives to constraints on $(p_1,p_2)$, and the solution reads
\bea
p_1=\ell-p_2,\quad p_2=\pm \frac{ian}{\sqrt{M^2-4a^2}}
\eea
Picking the first branch, we arrive at the final expression for the solution regular at the horizon:
\bea
\Phi=x^{\ell-p_2}(x-1)^{p_2} F\left[-\ell,-\ell+2p_2,1+2p_2,1-\frac{1}{x}\right],\quad 
p_2=\frac{ian}{\sqrt{M^2-4a^2}}
\eea
To extract the Love numbers we have to go to the ingoing Eddington--Finkelstein coordinates $(r,\psi)$, where $\psi$ is defined by 
\bea
d\psi=d\phi+\frac{a}{\Delta}dr=d\phi-\frac{a}{\mu}d\ln\left[\frac{2r-M+\mu}{2r-M-\mu}\right]=
d\phi-\frac{a}{\mu}d\ln\left[\frac{x}{x-1}\right],
\eea
and $\mu=\sqrt{M^2-4a^2}$. Then the wavefunction $\Psi$ defined by (\ref{WaveEqnTwo}) reads
\bea
\Psi&=&e^{in\phi}\Phi(r) X(\theta)=e^{in\psi}\left[\frac{x}{x-1}\right]^{\frac{ina}{\mu}}\Phi(r) X(\theta)\\
&=&e^{in\psi}X(\theta)\, \left\{x^{\ell} F\left[-\ell,-\ell+2p_2,1+2p_2,1-\frac{1}{x}\right]\right\}.
\nonumber
\eea
For integer $\ell$, the expression in the curly brackets is a polynomial of degree $\ell$ in the $r$ coordinate, so the expansion (\ref{PhiAsymp}) has no negative powers of $r$, thus all Love numbers vanish.

\section{Logarithmic running in odd dimensions}
\label{AppRun}

In this appendix we provide some details supporting statements (a)-(h) on page \pageref{PageAZ}. Specifically, we will study equation (\ref{PhiXeqn}), and it is convenient to denote its left--hand side by $F$
\bea\label{DefRhsF}
F=x^{-\ell/2-1}\Bigg[4x^2(x^n-M x+Ma^2)\Phi''+4x(n x^n-M x)\Phi'-\ell(2n+\ell-2)x^n\Phi\Bigg].
\eea
We begin with looking at the expansion (\ref{PsiAsymp}). 
Substituting it into function $F$, we find
\bea\label{FstartAA}
F=-\ell^2 M-4\sum s(n+\ell-s-1)c_s x^{n-s-1}-4b_3(\ell+n-1)x^{-\ell}+\OO(x^{-1}).
\eea
This implies that
\bea\label{cSeq0}
c_s=0,\quad\mbox{for}\quad s<n-1.
\eea
Other coefficients can be determined by solving equations order--by--order, and the first few terms are simple, for example\footnote{These answers are written for the generic case of $\ell>1$, and degenerate situations will be discussed below.}
\bea\label{cNforK0}
c_{n-1}=-\frac{\ell M}{4(n-1)},\quad c_n=\frac{\ell (\ell-2)a^2 M}{4(\ell-1)n}\,.
\eea
Using these relations as well as (\ref{cSeq0}), we find the expansion
\bea\label{tempAug17}
F&=&-4b_3(\ell+n-1)x^{-\ell}-4\sum_{s=2}^\infty (\ell-s)(s+n-1) c_{n+s-1}x^{-s} +
\frac{\ell (2+\ell-2n)^2M^2}{4(n-1)}x^{1-n}\nn
&&+
\OO(x^{-n})+\OO(x^{-\ell-1})+\ln x\left[\sum \alpha_q x^{-q}\right]
\eea
where coefficients $\alpha_q$ are some combinations of ${\hat c}_s$ from (\ref{PsiAsymp}). The last expression implies\footnote{Notice that the contribution with $s=\ell$ is absent from the sum, so for $n-1>\ell$ it is only the first term in (\ref{tempAug17}) that contains $x^{-\ell}$.} that coefficient $b_3$ vanishes for sufficiently large values of $n$:
\bea\label{Aug170one}
n-1>\ell>1\quad\Rightarrow\quad b_3=0\,.
\eea
We already observed this property for the static black holes in section \ref{SecStatic}, and now we see that it persists in the rotating case as well. The special values, $\ell=0,1$, can be analyzed in the same way\footnote{Recall that relations (\ref{cNforK0}) no longer hold, so one has to go 
back to (\ref{FstartAA}).}, and the results read
\bea\label{Aug170oneL01}
n>2:&&\ell=0\quad\Rightarrow\quad b_3=0,\qquad 
\ell=1\quad\Rightarrow\quad b_3=-\frac{a^2 M}{4n}\,;\\
n=2:&&\ell=0\quad\Rightarrow\quad b_3=0,\qquad 
\ell=1\quad\Rightarrow\quad b_3=\frac{M^2}{32}-\frac{a^2 M}{4n}\,.\nonumber
\eea
For $\ell=n-1>1$, equation $F=0$ with  (\ref{tempAug17}) gives
\bea\label{Aug17two}
\ell=n-1\quad\Rightarrow\quad b_3=\frac{(n-1)M^2}{32}\,.
\eea
We conclude that the logarithmic running found in the static case 
(see (\ref{StaticRunOne})--(\ref{StaticKhalf})) is not modified by the rotation parameter.

Given the special role played by $\ell=k(n-1)$ with integer $k$ already in the static case, it is convenient to write $\ell$ in terms of integers $(k,p)$ as
\bea\label{EllAsKapp}
\ell=k(n-1)+p,\quad p<n-1.
\eea 
Then equations (\ref{Aug170one})--(\ref{Aug170oneL01}) become the statements about the $k=0$ block. Furthermore,
inspired by the relation (\ref{cSeq0}), the expansion (\ref{PsiAsymp}) can be rearranged as
\bea\label{PsiAsympBlockApp}
\Phi&=&x^{\ell/2}\Bigg\{1+\sum_{s=1}^{k+1}f_s(y) x^{-s(n-1)}+
x^{-(k+1)(n-1)-p}\left[b_2\phi_2+
b_3 \phi_3 \ln x\right] \Bigg\}\,,
\eea
where $f_s$ is a polynomial in $y=x^{-1}$ whose degree is less than $(n-1)$. The first term in the curly bracket can be interpreted as $f_0=1$. Substituting (\ref{PsiAsympBlockApp}) into (\ref{DefRhsF}) and expanding $F$ in powers of $y$ for sufficiently large values of $n$, one concludes that polynomial $f_s$ has the degree $s$. Let us demonstrate this for $f_1$:
\bea
F&=&4x^2\frac{d^2 f_1}{dx^2}+\mu_1 x\frac{df_1}{dx}+\mu_2f_1+\mu_3+\frac{\mu_4}{x}
+\OO(x^{1-n}),\\
&&\mu_1=4\Big[p+(k-1)n+2-k\Big],\quad \mu_2=-4(n-1)\Big[k(n-1)+p\Big].\nonumber
\eea
Explicit expressions for constants $(\mu_3,\mu_4)$ are complicated, but they are not important for our analysis. Since the degree of the polynomial $f_1$ is less than $(n-1)$, the differential equation $F=0$ implies that
\bea
f_1(y)=f_{10}+f_{11}y.
\eea
Once $f_1(y)$ is fixed, the leading contributions to $F$ give a differential equation for $f_2$ and so on. Thus for any $s\le k$ we find
\bea\label{Aug17e3}
&&x^2\frac{d^2 f_s}{dx^2}+{\tilde\mu}_1 x\frac{df_s}{dx}+{\tilde\mu}_2f_s+
\sum_{q=0}^s {\tilde\mu}_{3+q} x^{-q}+\OO(x^{1-n})=0,\\
&&{\tilde\mu}_1=\Big[p+1+(k+1-2s)n_-\Big],\quad {\tilde\mu}_2=-sn_-\Big[(k-s+1)n_-+p\Big],\quad n_-=n-1,\nonumber
\eea
and the solution reads
\bea\label{SolnFs}
f_s(y)=\sum_{q=0}^s f_{sq}y^q\,,\quad s\le k\,.
\eea
Let us now quantify the earlier assertion that this construction works only ``for sufficiently large values of $n$''. The last term in equation (\ref{Aug17e3}) becomes negligible only if $n>s+1$, and if this condition is violated, then a separate analysis is required. Recalling that, by construction, the power of the polynomial $f_s$ should not exceed $(n-2)$, we conclude that the relevant solution of equation (\ref{Aug17e3}) is still given by (\ref{SolnFs}) but the sum goes only to $(n-2)$ rather than $s$. 

Let us first determine the values of $b_3$ for $n>k+1$. In this case all relevant polynomials $f_s$ with $s\le k$ appearing in (\ref{PsiAsympBlockApp}) are given by (\ref{SolnFs}). Equation for $f_{k+1}$ is a modified version of (\ref{Aug17e3}). For example, for $k=0$ we find
\bea\label{Aug18k0}
F&=&4x^2\frac{d^2 f_1}{dx^2}+\mu_1 x\frac{df_1}{dx}+\mu_2f_1+\mu_3+\frac{\mu_4}{x}
-4(n+p-1)b_3 x^{-p}+\OO(x^{1-n})
\\
&&\mu_1=4\Big[p-n+2\Big],\quad \mu_2=-4(n-1)p.\nonumber
\eea
Since by construction $f_1$ must be a polynomial in $y=x^{-1}$, it is only the $e_p y^p$ term in $f_1$ that can potentially balance the coefficient $b_3$ for $p>1$. However, coefficients 
$(\mu_1,\mu_2)$ ensure that all terms containing $e_p$ cancel in $F$. Therefore, we conclude that $b_3$ must vanish if $p>1$, in agreement with our earlier result (\ref{Aug170one}). 

To obtain new relations, we need the counterpart of (\ref{Aug18k0}) for $k>0$, and such extension of equation (\ref{Aug17e3}) reads
\bea\label{Aug18k1}
&&x^2\frac{d^2 f_{k+1}}{dx^2}+{\tilde\mu}_1 x\frac{df_{k+1}}{dx}+{\tilde\mu}_2f_{k+1}+
\sum_{q=0}^{k+1} {\tilde\mu}_{3+q} x^{-q}-\nu b_3 x^{-p}+\OO(x^{1-n})=0,\\
&&{\tilde\mu}_1=\Big[p+1-K\Big],\quad {\tilde\mu}_2=-Kp,\quad
\nu=\Big[p+K\Big],\quad K=(k+1)(n-1).\nonumber
\eea
Since $f_{k+1}$ is a polynomial in $y=x^{-1}$, and the term $y^p$ from this polynomial cancels between the first three contributions to the last equation, we conclude that $b_3$ vanishes for sufficiently large values of $p$:
\bea\label{B3triangle}
p>k+1\quad\Rightarrow\quad b_3=0.
\eea
To determine $b_3$ for other values of $p$, one needs to evaluate all coefficients $f_{sq}$ in (\ref{SolnFs}) and calculate the resulting ${\tilde\mu}_{3+q}$ in (\ref{Aug18k1}). Let us summarize the results:
\begin{enumerate}[1.]
\item As we already saw, $b_3$ vanishes for $p>k+1$.
\item The result for $k=0$ is given by (\ref{Aug170one}), (\ref{Aug170oneL01}).
\item For $k=1$ we find
\bea\label{RunK1gen}
&&p=0\ \Rightarrow\ b_3=\frac{(n-1)M^2}{32}\,;\qquad
p=1\ \Rightarrow\ b_3=-\frac{(n-2)(naM)^2}{8(2n-1)(n-1)}\,;\nn
&&p=2\ \Rightarrow\ b_3=\frac{(n^2-1)^2a^4M^2}{32n^2}\,.
\eea
\item For $k=2$ we find
\bea\label{RunK2gen}
&&p=0,2\ \Rightarrow\ b_3=0\,;\qquad
p=1\ \Rightarrow\ b_3=-\frac{(n-3)(2n-1)^2a^2 M^3}{64n(n-1)^2(3n-2)}\,;\nn
&&p=3\ \Rightarrow\ b_3=-\frac{(4n^2-1)^2a^6M^3}{768n^5}\,.
\eea
\item For $k=3$ we find
\bea\label{RunK3gen}
&&p=0\ \Rightarrow\ b_3=\frac{9(n-1)M^4}{4096}\,;\nn
&&p=1\ \Rightarrow\ b_3=-\frac{n(n-2)(3n-2)^2(4-5n+3n^2)a^2 M^4}{384(n-1)^3(2n-1)(4n-3)}\,;\nn
&&p=2\ \Rightarrow\ b_3=\mbox{$\frac{(3n-1)^2(36 - 153 n + 215 n^2 - 143 n^3 + 209 n^4 - 288 n^5 + 108 n^6)a^4M^4}{1024n^2(2n-1)^3(3n-2)^2}$} \,\\
&&p=3\ \Rightarrow\ b_3=-\frac{27n^4(n-2)(3n-2)a^6M^4}{128(n-1)(2n-1)(3n-1)^2(4n-1)}\,;\nn
&&p=4\ \Rightarrow\ b_3=-\frac{(n^2-1)^2(9n^2-1)^2a^8M^4}{36864 n^7}\,\nonumber
\eea
\item These calculations can be continued for all values of $k$, and the pattern is clear: $b_3$ has the form
\bea\label{B3pk}
b_3=B_3 a^{2p} M^{k+1}\,,
\eea
where $B_3$ is a complicated constant that depends only on $(n,k,p)$. In particular, this implies that for $p=0$, the running coefficient $b_3$ does not depend on the rotation parameter, so it is given by the static results evaluated in section \ref{SecStatic}. If $p\ne 0$, then $B_3$ vanishes in the static case, but it has a non--zero value for rotating black holes with the single exception of 
$k=p=2$.
\end{enumerate}
So far we assumed that $n>k+1$. If this condition is violated, then a special treatment is required, as we already saw in (\ref{Aug170oneL01}), and here we present additional examples:
\begin{itemize}
\item {$k=0$}\\
In this case, only $n=\{0,1\}$ would be special, but we are focusing on $n\ge 2$.
\item {$k=1$}\\
Only $n=2$ is special, then $\ell=1$, so we recover the result from the last line of equation  (\ref{Aug170oneL01}):
\bea\label{RunK1spec}
n=2, \ \ell=1\quad\Rightarrow\quad b_3=\frac{M^2}{32}-\frac{a^2 M}{8}\,.
\eea
\item {$k=2$}\\
There are two special values, $n=\{2,3\}$ and the results are:
\bea\label{RunK2spec}
n=2:&& \ \ell=2\quad\Rightarrow\quad b_3=0\,;\\
n=3:&& \ \ell=4\quad\Rightarrow\quad b_3=\frac{2a^4 M^2}{27}\,,\quad 
\ell=5\quad\Rightarrow\quad b_3=0\,.\nonumber
\eea
\item {$k=3$}\\
There are three special values, $n=\{2,3,4\}$ and the results are:
\bea\label{RunK3spec}
n=2:&&\ell=3\ \Rightarrow\ b_3=\frac{9}{2^{12}}M^2 [M-4a^2]^2\,;\\
n=3:&& \ell=6\ \Rightarrow\ b_3=\frac{9M^4}{2^{11}}\,,\quad
\ell=7\ \Rightarrow\ b_3=-\frac{49a^2 M^4}{2^8\cdot 3^6\cdot 5}[324+\frac{125 a^4}{M}]\,;
\nn
n=4:&& \ell=9\ \Rightarrow\ b_3=\frac{27M^4}{2^{18}}[64-\frac{49a^6}{M}]\,,\quad
\ell=10\ \Rightarrow\ b_3=-\frac{200 a^2 M^4}{3^4\cdot 7\cdot 13}\,,
\nn
&&\ell=11\ \Rightarrow\ b_3=\frac{11^2\cdot 23^3\,(a M)^4}{2^{12}\cdot 5^2\cdot 7^3}\,.
\nonumber
\eea
\item {$k=4$}\\
There are four special values, $n=\{2,3,4,5\}$. To avoid writing complicated formulas, we 
focus on the $p=0$ case, which in this situation gives zero according to the formula (\ref{B3pk}). By construction, this formula breaks down for $n\le k+1=5$, and the special solutions are
\bea\label{RunK4spec}
&&(n,\ell)=(2,4)\ \Rightarrow\ b_3=0;\quad (n,\ell)=(3,8)\ \Rightarrow\ b_3=\frac{2\cdot 3^4\, (aM)^4}{5^3\cdot 7^2}\,;\\
&&(n,\ell)=(4,12)\ \Rightarrow\ b_3=-\frac{24 a^6 M^4}{7\cdot 11^2}\,;
\quad (n,\ell)=(5,16)\ \Rightarrow\ b_3=\frac{28^2\, a^8M^4}{5^7}\,.
\nonumber
\eea
\end{itemize}
For $n>2$, these expressions have the structure
\bea\label{B3pkSpec}
b_3=B_3 a^{2p} M^{k+1}+{\hat B}_3 a^{2(p+n-1)} M^{k}\,,
\eea
where the values of $B_3$ can still be obtained by formally using the expressions (\ref{RunK1gen})--(\ref{RunK3gen}), even though they were derived assuming that $n>k+1$. The values of ${\hat B}_3$ are nonzero for 
\bea
(k,n,p)\ =\ (2,3,0),\ (3,3,1),\ (3,4,0),\ (4,3,0),\ (4,3,1),\ (4,4,0),\ (4,4,1),\ (4,5,0).\nonumber
\eea

\bigskip
\noindent
To summarize, in this appendix we provided some technical details supporting statements (a)-(h) on page \pageref{PageAZ}. Specifically, we analyzed the parameter $b_3$, which describes the logarithmic running, as a function of the number of dimensions and the angular momentum of an excitation. Writing the latter as (\ref{EllAsKapp}), we identified the patterns corresponding to various values of $(p,n)$. In particular, we showed 
that for $k\le n-1$, the running coefficient $b_3$ is given by (\ref{B3pk}), and we computed the numerical coefficients $B_3$. For larger values of $k$, the coefficient $b_3$ by (\ref{B3pkSpec}).

\section{Wave equation in five dimensions}
\label{App5d}

In this appendix we review the evaluation of the Love numbers for five--dimensional black holes carried out in \cite{RotLN5d,RSST}\footnote{See also \cite{LoveCharge4} for the extension to the charged solutions.}. These articles introduced convenient variable which allowed the authors to solve the wave equation in terms of hypergeometric functions, but unfortunately as we will see, this procedure breaks down when two rotation parameters become equal. After reviewing the known results, we propose a different convenient variable, and an alternative solution of the wave equation in terms of hypergeometric functions, which admits a straightforward reduction to the case of equal rotations. 

\bigskip

We begin with reviewing separability of the equation for a scalar field in the background of the rotating black hole in five dimensions. The relevant geometry is obtained by setting $n=2$ in (\ref{MPodd}):
\bea\label{MPodd5d}
\hskip -0.3cm
ds^2=-dt^2+\frac{M}{\Sigma}\Big(dt+\sum_{i=1}^{2} a_i\mu_i^2 d\phi_i\Big)^2+
\frac{r^2\Sigma dr^2}{\Delta}
+\sum_{i=1}^{2}(r^2+a_i^2)\Big(d\mu_i^2+\mu_i^2 d\phi_i^2\Big),
\eea
To agree with the standard notation used in five dimensions, we introduced several convenient quantities\footnote{We also use a shorthand notation for the trigonometric functions: $s_\theta=\sin\theta$, $c_\theta=\cos\theta$.} :
\bea\label{Sigma5d}
&&(\mu_1,\mu_2)=(s_\theta,c_\theta),\quad (a_1,a_2)=(a,b),\quad
\Delta=R-M r^2,\quad R=R_aR_b,\nn
&&\Sigma=r^2+\Theta^2,\quad 
\Theta=\left[(ac_\theta)^2+(bs_\theta)^2\right]^{\frac{1}{2}}\,,\
R_a=(r^2+a^2),\ R_b=(r^2+b^2).
\eea
The special frames (\ref{AllFramesMPOdd}) are given by 
\bea\label{AllFramesBH5}
e_t&=&-\frac{R}{r\sqrt{\Delta\Sigma}}\left[ \d_t
-\frac{a}{r^2+a^2}\d_{\phi}-\frac{b}{r^2+b^2}\d_{\psi}\right],\quad e_r=\sqrt{\frac{\Delta}{r^2\Sigma}}\d_r,\quad
e_{\theta}=\frac{1}{\sqrt{\Sigma}}\d_{\theta}\nn
\\
e_1&=&-\frac{s_\theta c_\theta}{\Theta\sqrt{\Sigma}}\left[(a^2-b^2)\d_t-\frac{a}{s_\theta^2}\d_{\phi}+
\frac{b}{c_\theta^2}\d_{\psi}
\right],\quad 
e_\psi=-\frac{ab}{r\Theta}\left[\d_t-\frac{1}{a}\d_{\phi}-\frac{1}{b}\d_{\psi}\right].\nonumber
\eea
The radial part of the wave equation (\ref{MPodeOdd}) in five dimensions reads
\bea\label{SeparWave5d}
r\frac{d}{dr}\left[\frac{\Delta}{r}\frac{d\Phi}{dr}\right]+\frac{R^2}{\Delta}
\left[\omega-
\sum_{k=1}^2\frac{a_k n_k}{r^2+a_k^2}\right]^2\Phi-\Omega^2\Phi-\Lambda r^2\Phi=0,
\eea
where $\Lambda$ is an eigenvalue of the angular equation, and $\Omega$ is defined by
\bea
\Omega=(ab)\left[\omega-\frac{n_\phi}{a}-\frac{n_\psi}{b}\right].\nonumber
\eea
For the subsequent analysis, it is convenient to rewrite equation (\ref{SeparWave5d}) as
\bea\label{eqn5aOne}
r\frac{d}{dr}\left[\frac{\Delta}{r}\frac{d\Phi}{dr}\right]+\frac{F r^2}{\Delta} \Phi-\Lambda r^2\Phi=0
\eea
where function $F(r)$ is defined by
\bea
F=M (bn_\phi+an_\psi-a b \omega)^2 +(a^2-b^2)\left\{\left[n_\phi-\frac{a \omega R_a}{a^2-b^2}\right]^2 R_b-\left[n_\psi-\frac{b \omega R_b}{b^2-a^2}\right]^2 R_a\right\}.\nonumber
\eea
We are primarily interested in the $\omega=0$ case, when function $F$ simplifies to
\bea\label{eqn5aTwo}
F=M (bn_\phi+an_\psi)^2 +(a^2-b^2)\Big[(n_\phi)^2 R_b-(n_\psi)^2 R_a\Big].
\eea
Equations (\ref{eqn5aOne}), (\ref{eqn5aTwo}) can be used to extract the Love numbers for the five--dimensional black holes \cite{RSST}, but before doing that we should determine the eigenvalues $\Lambda$ entering (\ref{eqn5aOne}). These values are obtained by solving the angular counterpart of the radial equation (\ref{SeparWave5d}), which reads
\bea\label{SeparWave5dAng}
\frac{1}{s_{2\theta}}\frac{d}{d\theta}\left[s_{2\theta}\frac{dX}{d\theta}\right]-
\frac{(s_\theta c_\theta)^2}{\Theta^2}\left[(a^2-b^2)\omega-\frac{a n_\phi}{s^2_\theta}+\frac{b n_\psi}{c^2_\theta}\right]^2 X-\frac{\Omega^2}{\Theta^2}X+\Lambda X=0
\eea
As in the radial case, this equation can be rewritten in the form similar to (\ref{eqn5aOne}),
\bea\label{eqn5aTwoA}
\frac{1}{s_{2\theta}}\frac{d}{d\theta}\left[s_{2\theta}\frac{dX}{d\theta}\right]-
\frac{F_\theta}{\Theta^2} X+\Lambda X=0,
\eea
where function $F_\theta$ is defined by
\bea
F_\theta=\frac{(ac_\theta)^2+(b s_\theta)^2}{(s_\theta c_\theta)^2}
\Big[c_\theta^2(n-a\omega s_\theta^2)^2+s_\theta^2(m-b\omega c_\theta^2)^2\Big]
\eea
For the time--independent modes of the scalar field, i.e., for $\omega=0$, rotation parameters $(a,b)$ 
disappear in equation (\ref{eqn5aTwoA}), so one gets the eigenvalue problem for the standard spherical harmonics, which leads to 
\bea\label{5dLam}
\Lambda=\ell(\ell+2)\quad\mbox{with integer}\quad \ell.
\eea 
To summarize, the radial profile of the scalar field satisfies equations (\ref{eqn5aOne}) with function $F$ given by  (\ref{eqn5aTwoA}) and constant $\Lambda$ given by (\ref{5dLam}). 

\subsection{Review of the known construction}
\label{App5dKnown}

The system (\ref{eqn5aOne}),  (\ref{eqn5aTwo}),  (\ref{5dLam}) was analyzed in \cite{RSST} under the assumption 
that rotation parameters $(a,b)$ are not equal. Let us begin with reviewing these results before presenting an alternative construction that would work for equal rotations as well. To simplify equation (\ref{eqn5aOne}), the authors of \cite{RSST} proposed a new coordinate $x$ defined by
\bea\label{RsqRodr}
r^2 = \frac{(r_+^2-r_-^2)x + r_{+}^2+r_{-}^2}{2}\,,
\eea
where $r_\pm$ are two positive solutions of equation $\Delta=0$. Since $\Delta$ is an even polynomial of degree
four in $r$  (see (\ref{Sigma5d})), the roots of equation $\Delta=0$ are $r=(\pm r_+,\pm r_-)$, and it terms of variable $x$ one finds
\bea
\Delta=\frac{1}{4}(r_+^2-r_-^2)^2(x^2-1).\nonumber
\eea
The wave equation (\ref{eqn5aOne}) becomes \cite{RSST}
\bea\label{eqn5dX}
\frac{d}{dx}\left[(x^2-1)\frac{d\Phi}{dx}\right]+
2\left[\frac{({m_L}+{m_R})^2}{x-1}-\frac{({m_L}-{m_R})^2}{x+1}\right]\Phi- \frac{1}{4}\ell(\ell+2)\Phi=0.
\eea
Here constants $(m_L,m_R)$ are defined by
\bea
{m_L}=\frac{a-b}{r_+  + r_-}\frac{n_\phi-n_\psi}{4} \quad {m_R}=\frac{a+b}{r_+  - r_-}\frac{n_\phi+n_\psi}{4}\,.
\eea
The authors of \cite{RSST} proposed another change of variables,
\bea
\Phi(z)=z^{\ell/2+1}(1-z)^{i({m_L}+{m_R})} u(z),\quad z=\frac{2}{1+x} 
\eea
which puts equation (\ref{eqn5dX}) in the standard hypergeometric form 
\bea
z(1-z)u''+\Big[\mathfrak{c}-(\mathfrak{a}+\mathfrak{b}+1)z\Big]u'-
\mathfrak{a}\,\mathfrak{b}\,u=0
\eea
with parameters $(\mathfrak{a},\mathfrak{b},\mathfrak{c})$ given by 
\bea
\mathfrak{a}=1+\frac{\ell}{2}+2im_L,\quad \mathfrak{b}=1+\frac{\ell}{2}+2im_R,\quad \mathfrak{c}=1+{\ell}\,.
\eea
The unique solution regular at the horizon, which is located at $r=r_+$ or $x=z=1$, is
\bea\label{eqn5dSolnApp}
\Phi=z^{\ell/2+1}(1-z)^{i({m_L}+{m_R})}F\Big[\mathfrak{a},\mathfrak{b},1+\mathfrak{a}+\mathfrak{b}-\mathfrak{c},1-z\Big]
\eea
As expected from the general analysis presented in section \ref{SecLogRunSec}, at large values of $r$, this solution behaves as 
(\ref{PsiAsympGen}) with $d=5$:
\bea\label{PsiAsympD5App}
\Phi=b_1 r^\ell\Bigg\{\sum_{s=0}^{k+1}\frac{M^s}{r^{2s}} f_s(r)+
r^{-2-2\ell}\left[b_2\phi_2+
b_3 \phi_3 \ln r^2\right] \Bigg\}\,.
\eea
with $f_0(0)=1$. The values of $(b_2,b_3)$ follow from the expansion of (\ref{eqn5dSolnApp}) near $z=0$, and they can be found in \cite{RSST}.

\subsection{New parameterization of scalar excitations}
\label{App5dNew}

Let us go back to the system (\ref{eqn5aOne}),  (\ref{eqn5aTwo}),  (\ref{5dLam}). We begin with looking at solutions without dependence on cyclic coordinates, then the radial equation becomes
\bea\label{eqn5aOneNoCyc}
r\frac{d}{dr}\left[\frac{\Delta}{r}\frac{d\Phi}{dr}\right]-\ell(\ell+2) r^2\Phi=0
\eea
For $a=b=0$, the solution regular at the horizon is given by a special case of equation (\ref{PhiHyper}):
\bea\label{eqn5dStat}
\Phi=c_1 r^{\ell}F\left[-\frac{\ell}{2},-\frac{\ell}{2},1,1-\frac{M}{r^{2}}\right]\,.
\eea
Interestingly, this solution survives in the rotating case with a minor modification. Specifically, introducing a new coordinate 
\bea\label{Xhyper1}
x=a_1 r^2+a_2
\eea
and requiring (\ref{eqn5aOneNoCyc}) to reduce to the hypergeometric equation, we find the values of the parameters 
$(a_1,a_2)$:
\bea\label{Xhyper2}
a_2 = \frac{1}{2} \left[1 + ( a^2 + b^2-M) a_1\right], \quad
a_1 = \frac{1}{\sqrt{(M-a^2-b^2)^2-4(a b)^2}}\,.
\eea
The solution of equation (\ref{eqn5aOneNoCyc}) reads
\bea\label{5dHyperNC}
\Phi=x^{\frac{\ell}{2}} F[-\frac{\ell}{2}, -\frac{\ell}{2}, 1, 1 - \frac{1}{x}]\,.
\eea
Motivated by this expression, we write the solution of the system (\ref{eqn5aOne}),  (\ref{eqn5aTwo}),  (\ref{5dLam}) in the form
\bea
\Phi=x^{p_1}(x-1)^{p_2}g\left[1-\frac{1}{x}\right].
\eea
The differential equation reads
\bea\label{5dPreHyper}
z(1-z)g''+\left[1+2 p_2+(2p_1-1)z\right]g'+\frac{G(z)}{z(1-z)}g=0
\eea
The explicit form of polynomial $G(z)$, which depends on $(p_1,p_2)$, is rather complicated. To get the hypergeometric equation, we must impose two conditions,
\bea
G(0)=0,\quad G(1)=0,\nonumber
\eea
which lead to relations\footnote{There is also an alternative option $p_1=-\frac{\ell}{2}-1-p_2$ and s slightly different expression for $p_2$, and it will be discussed below.}
\bea\label{5dP1P2}
p_1&=&\frac{\ell}{2}-p_2,\\ 
(p_2)^2&=&\frac{a_1(p^2-q^2)(b^2-a^2)}{8}-\frac{a_1^2 M abpq}{2}
+\frac{a_1^2(p^2+q^2)}{8}[(a^2-b^2)^2-M(a^2+b^2)]\,.\nonumber
\eea
To avoid unnecessary clutter in the above expression and subsequent formulas, we labeled the angular quantum numbers $(n_\phi,n_\psi)$ as $(p,q)$:
\bea\label{pqQuantNum}
p=n_\phi,\quad q=n_\psi\,.
\eea 
Solution (\ref{5dP1P2}) gives two branches:
\bea
p_2=\pm \frac{1}{4}\left[\frac{(a+b)(p+q)}{\sqrt{M-(a+b)^2}}+\frac{(a-b)(p-q)}{\sqrt{M-(a-b)^2}}\right]\,.
\eea
With parameters $(p_1,p_2)$ given by (\ref{5dP1P2}), equation (\ref{5dPreHyper}) reads
\bea\label{5dHyper}
z(1-z)g''+\left[(1+2 p_2)(1-z)+\ell z\right]g'-\frac{\ell(\ell-4p_2)-a_1(a^2-b^2)(p^2-q^2)}{4}g=0.
\eea
The solution regular at $z=0$ reads
\bea\label{5dHyperSoln}
&&\Phi=x^{\ell/2}\left[1-\frac{1}{x}\right]^{p_2}g\left[1-\frac{1}{x}\right],\quad 
g(z)=F\left[A-B,A+B,1+2p_2,z\right],\nn 
&&A=p_2-\frac{\ell}{2},\quad B=\left[(p_2)^2+\frac{a_1}{4}(a^2-b^2)(p^2-q^2)\right]^{\frac{1}{2}}\,,
\eea
where parameter $p_2$ is given by (\ref{5dP1P2}). 

The relation (\ref{5dHyperSoln}) constitutes the final result of this subsection, and it gives  the regular solution for all values of $(a,b,p,q)$ which reduces to (\ref{eqn5dStat}) when the angular quantum numbers are turned off. It is instructive to consider some special cases.
\begin{itemize}
\item {\bf The modes independent of cyclic coordinates.}\\
Setting $p=q=0$ in (\ref{5dHyperSoln}), (\ref{5dP1P2}), we find
\bea\label{5dAnsNC}
p_2=0,\quad \Phi=x^{\ell/2}F\left[-\frac{\ell}{2},-\frac{\ell}{2},1;1-\frac{1}{x}\right].
\eea
This recovers the simple expression (\ref{5dHyperNC}) found earlier.
\item {\bf The solutions with $b=0$.}\\
Removing one of the rotations in (\ref{5dHyperSoln}), (\ref{5dP1P2}), we find the parameters of the solution,
\bea
(p_2)^2&=&\frac{a_1(q^2-p^2)a^2}{8}
+\frac{a_1^2(p^2+q^2)}{8}[a^4-M a^2]=-\frac{a^2 p^2}{4(M-a^2)},\nn
a_1&=&\frac{1}{M-a^2}\,,\quad B=-\frac{aq}{2\sqrt{M-a^2}}
\eea
as well as the wavefunction 
\bea
\Phi=x^{\ell/2}\left[1-\frac{1}{x}\right]^{p_2} F\left[-\frac{\ell}{2}+p_2-B,-\frac{\ell}{2}+p_2+B,1+2p_2,1-\frac{1}{x}\right]
\eea
Note that $B$ and $p_2$ are imaginary. If $a$ is set to zero as well, then the wavefunction reduces to (\ref{5dAnsNC}), so information about angular quantum numbers disappears, as expected in the spherically symmetric case. 
\item {\bf The solutions with $a=b$.}\\
Setting two rotations in (\ref{5dHyperSoln}), (\ref{5dP1P2}) to be equal, we find the parameters of the solution,
\bea
(p_2)^2=-\frac{M a_1^2 a^2(p+q)^2}{4},\quad
a_1=\frac{1}{\sqrt{M(M-4a^2)}}\,,\quad B=p_2\,,
\eea
as well as the wavefunction 
\bea
\Phi=x^{\ell/2}\left[1-\frac{1}{x}\right]^{p_2} F\left[-\frac{\ell}{2},-\frac{\ell}{2}+2p_2,1+2p_2,1-\frac{1}{x}\right]
\eea
As before, $p_2$ is imaginary.
\end{itemize}

\subsection{Extremal black holes}
\label{App5dExtrem}

Note that changes of variables (\ref{RsqRodr}) and (\ref{Xhyper1})--(\ref{Xhyper2}) break down in the extremal case. Indeed, when $M=(a+b)^2$, we find\footnote{Without loss of generality we assumed that rotational parameters $a$ and $b$ are positive. In the case of mixed signs, $a<0<b$, the extremal black hole corresponds to $M=(b-a)^2$ and $\Delta=(r^2+ab)^2$. The rest of the discussion of this subsection proceeds with minor modifications.}
\bea
\Delta=(r^2-ab)^2,
\eea
so $r_+=r_-=\sqrt{ab}$ in (\ref{RsqRodr}), and $a_1=\infty$ in (\ref{Xhyper1})--(\ref{Xhyper2}). Therefore, we have to go back to the extremal limit of the system (\ref{eqn5aOne})--(\ref{eqn5aTwo}). The resulting equation reads
\bea\label{eqn5aOneExtr}
&&r\frac{d}{dr}\left[\frac{(r^2-ab)^2}{r}\frac{d\Phi}{dr}\right]+\frac{F r^2}{(r^2-ab)^2} \Phi-\Lambda r^2\Phi=0,
\\
&&F=(a+b)^2(bp+aq)^2 +(a^2-b^2)\Big[p^2 R_b-q^2 R_a\Big].\nonumber
\eea
As in previous subsection, we introduced labels $(p,q)$ for the angular quantum numbers via (\ref{pqQuantNum}). Transitioning to a natural variable $z$ and a new function $Z(z)$ defined by 
\bea
z=\frac{b_1}{r^2-ab},\quad \Phi(r)=\sqrt{z}Z(z),
\eea
we arrive at the differential equation
\bea\label{eqn5dBessel}
4z^2 Z''+4z Z'-(\ell+1)^2Z+\frac{(a^2-b^2)(p^2-q^2)}{b_1}z Z+\Xi^2 z^2 Z=0.
\eea
Here constant $\Xi$ is defined by
\bea
\Xi^2=\frac{(a+b)^2}{(b_1)^2}\left[(a^2+b^2)(p^2+q^2)-ab(p-q)^2\right]
\eea
For $a=b$ or $p=\pm q$, equation (\ref{eqn5dBessel}) can be solved in terms of Bessel functions, and the result reads
\bea
Z=c_1 J_{(\ell+1)/2}\left[\frac{{\Xi}\,z}{2}\right]+c_2 Y_{(\ell+1)/2}\left[\frac{{\Xi}\,z}{2}\right]
\eea
As one approaches the horizon, the coordinate $z$ goes to infinity, and the leading asymptotics of the last expression is given by
\bea
Z=\left[\frac{4}{\Xi z}\right]^{\frac{1}{2}}\left[c_1\cos\zeta+c_2\sin\zeta\right],\quad \zeta=z-\frac{\pi (\ell+1)}{4}-\frac{\pi}{4}
\eea
The ingoing boundary at the horizon corresponds to $c_2=-ic_1$. 

For generic values of $(a,b,p,q)$, the solution of equation (\ref{eqn5dBessel}) can be written in terms of the Kummer's confluent hypergeometric function $\mathfrak{M}$: 
\bea
Z&=&e^{-i\Xi\, z/2}\left\{c_1 z^{\frac{1+\ell}{2}}
\mathfrak{M}\left[iA+1+\frac{\ell}{2},2+\ell,i\Xi z\right]
+
c_2 z^{-\frac{1+\ell}{2}}\mathfrak{M}\left[iA-\frac{\ell}{2},-\ell,i\Xi z\right]
\right\},\nn
A&=&\frac{(a^2-b^2)(p^2-q^2)}{4b_1\Xi}\,.
\eea
We recall that function $\mathfrak{M}$ is defined by the expansion
\bea
\mathfrak{M}(\aa,\bb,x)&=&\sum_{n=0}^{\infty}\frac{(\aa)_n}{(\bb)_n}\frac{x^n}{n!}
\eea
Ingoing boundary conditions at the horizon fix the ratio $c_2/c_1$, and the resulting solution can be written it terms of the Tricomi function $U$:
\bea\label{Tricomi}
Z=e^{-i\Xi\, z/2}z^{\frac{1+\ell}{2}}U\left[iA+1+\frac{\ell}{2},2+\ell,i\Xi z\right]
\eea
The values of $(c_1,c_2)$ can be extracted from the relation
\bea
U(\aa,\bb,x)&=&\frac{\Gamma(1-\bb)}{\Gamma(\aa-\bb+1)}\mathfrak{M}(\aa,\bb,x)+
\frac{\Gamma(\bb-1)}{\Gamma(\aa)}x^{1-\bb}\mathfrak{M}(\aa-\bb+1,2-\bb,x).
\eea
Implications of the solution (\ref{Tricomi}) are discussed in section \ref{SecN1d5}.

\section{Technical details for extremal black holes}

In this appendix, which supplements the discussion presented in section \ref{SecExtremal}, we provide some details pertaining to evaluation of Love numbers for extremal black holes in seven and nine dimensions. Specifically, in the seven-dimensional case, we carry out the analysis of equation (\ref{PellMain}) leading to the special set of angular momenta (\ref{EllFromPell}) which avoid running, and we also quote the values of the rational coefficients $({\bar b}_2,{\hat b}_2)$ appearing in the Love numbers (\ref{PsiRunExtrm}). In the nine--dimensional case, we analyze the  radial equation and demonstrate that its solutions can be reduces to the hypergeometric function only for $\ell=(-2,-4)$ (in addition to the trivial case of $\ell=(0,-6)$) and provide some details leading to the resulting equation (\ref{PhiXeqnAextr9d}). In section \ref{AppExtrBC} we address a different but related problem of excitations with nontrivial quantum numbers in cyclic directions. The resulting analysis justifies the conclusions presented in section \ref{SecN1gen}.

\subsection{Perturbations without angular dependence}
\label{AppExtremal}

As we discussed in section \ref{SecExtremal}, there is no logarithmic running in the asymptotic expansion (\ref{PsiAsympExtrm}) if the relation (\ref{PellMain}) is satisfied with a non--negative integer $k$. Let us construct the solutions of this equation with integer values of $\ell$. We begin with rewriting equation (\ref{PellMain})  in a more explicit form:
\bea\label{PellApp}
\sqrt{\frac{(\ell+1)(\ell+3)}{3}}-\ell-1=-2k
\eea
Rearrangement of terms leads to a quadratic equation for $\ell$,
\bea\label{EllEqnPrePell}
\ell^2 + (1 - 6k)\ell + \frac{3(1 - 2k)^2 - 3}{2} = 0,
\eea
which has integer solutions,
\bea\label{EllPellTemp}
\ell=\frac{6k-1\pm p}{2}\,,
\eea
if and only if its discriminant is a square of an integer $p$:
\bea
3\left(2k + 1\right)^2 - 2=p^2\,.\nonumber
\eea
This is a special case of the Pell's equation \cite{Pell}, and to proceed we rewrite it in the standard form
\bea\label{PellEqn}
p^2-3q^2=-2.
\eea
We are interested in solutions with integer values of $p$ and odd values of $q$. Clearly, $p=q=1$ is an example of such a solution. To find other solutions, we look at the ring of integers
$\mathbb{Z}[\sqrt{3}]$, whose elements have the form $\xi=p+q\sqrt{3}$. Given the multiplication rule,
\bea
\xi_1\cdot \xi_2=(p_1p_2+3 q_1q_2)+(p_1q_2+p_2q_1)\sqrt{3}\,,
\eea 
it is natural to define the norm of such numbers by
\bea
N[p+q\sqrt{3}]=p^2-3q^2.
\eea
This norm has an important multiplicative property,
\bea
N[\xi_1\cdot\xi_2]=N[\xi_1]\,N[\xi_2],
\eea
and equation (\ref{PellEqn}) implies that we are looking for the elements of $\mathbb{Z}[\sqrt{3}]$ whose norm is equal to minus two. We already have one candidate, $\eta=1+\sqrt{3}$, and it it clear that the most general solution would have the form
\bea
\xi=\eta\cdot \zeta,
\eea
where $\zeta$ is an element of $\mathbb{Z}[\sqrt{3}]$ that has unit norm. It is easy to see that up to a sign, any such element can be written as\footnote{Formally, this is guaranteed by the Dirichlet's unit theorem, but the result can also be obtained directly by observing that for any positive element $\zeta$ with a unit norm, there exists an integer $t$ such that $
1\le \zeta {\bar\zeta}^t<{\bar\zeta}$. Then a direct calculation shows that 
$\zeta {\bar\zeta}^t=1$.}
\bea\label{DefZetaBar}
\zeta={\bar \zeta}^s,\quad {\bar\zeta}=2+\sqrt{3}\,,
\eea
where $s$ is an arbitrary integer. To summarize, at this point we have found that\footnote{The second relation is obtained by taking a conjugate in $\mathbb{Z}[\sqrt{3}]$, which amounts to flipping the sign of $\sqrt{3}$ everywhere.}
\bea
p+(2k+1)\sqrt{3}=(1+\sqrt{3})\cdot {\bar\zeta}^s\,,\quad 
p-(2k+1)\sqrt{3}=(1-\sqrt{3})\cdot {\bar\zeta}^{-s}\,.
\eea
Solving these equations, for $(p,k)$ and substituting the result into equation (\ref{EllPellTemp}), we find the positive values of $\ell$:
\bea
\ell=\frac{1}{2}\left[(2-\sqrt{3}){\bar\zeta}^{-s}+(2+\sqrt{3}){\bar\zeta}^{s}\right]-2
\nonumber
\eea
Recalling the definition of ${\bar\zeta}$ from (\ref{DefZetaBar}), we arrive at the final expression for $\ell$:
\bea
\ell=\frac{1}{2}\left[{\bar\zeta}^{-s-1}+{\bar\zeta}^{s+1}\right]-2\,.
\eea
This completes the derivation of equation (\ref{EllFromPell}). 

\bigskip

\noindent
Let us now present the data for the rational numbers 
$({\bar b}_2,{\hat b}_2)$ in equation (\ref{PsiRunExtrm}). Unfotunately, we were not able to write these numbers in closed form for arbitrary values of $\ell$, but they can be computed on the case--by--case basis. To keep the expressions compact, we define
\bea\label{ExtremBdef}
B(\ell)={\bar b}_2+{\hat b}_2 s.
\eea
Then direct evaluation gives
\bea
&&B(1)=-\frac{201-81s}{64}\,,\quad
B(2)=\frac{3^5\cdot 41}{2^{10}}\,,\quad
B(3)=-\frac{3^9(18+5s)}{2^{10}\cdot 5^2}\,,\nn
&&B(4)=\frac{3^3 (50+63s)}{5\cdot 2^6}\,,\quad
B(5)=\frac{5\cdot 3^8}{2^{17}}\,,\quad
B(6)=\frac{3^{12} \left[(5\cdot 19)^2-3\cdot 2^9 (s-4)\right]}{5\cdot 7\cdot 2^{22}}\,,\nn
&&B(7)=\frac{21\cdot 17\cdot\left[77\cdot 353-3^5 2^{10} (s+49)\right]}{25\cdot 2^{20}}\,,
\nn
&&B(8)=\frac{11\cdot 3^{13}\cdot \left[2^{11}+35\cdot 3^2 (s+1)\right]}{
2^{10}\cdot 5^4\cdot 7^3}\,,
\\
&&B(9)=-\frac{3^{14} \left[7\,702\,357\,493+7\cdot 11\cdot  197\cdot  2^6 \cdot 3^8 s\right]}{2^{30}\cdot 5^2\cdot 7^3\cdot 11^2}\,,\nn
&&B(10)=\frac{3\cdot 17\cdot 25\cdot 29\cdot \left(121\cdot 8\,660\,077-
13\cdot 89\cdot 3^4\cdot 2^{11}s\right)}{7\cdot 11\cdot 2^{33}}\,.\nonumber
\eea
Substitution of these formulas into (\ref{ExtremBdef}) gives the coefficients $({\bar b}_2,{\hat b}_2)$, which in tern leads to the final expressions for the Love numbers in (\ref{PsiRunExtrm}).

\bigskip

\noindent
We conclude this appendix by analyzing the wave equation on the background of an extremal nine--dimensional black hole and demonstrating that the problem reduces to the hypergeometric equation only for particular values of the separation constant. 

Our starting point is equation (\ref{PhiXeqnAextr}) for $n=4$:
\bea\label{Extrm9dApp}
\frac{4}{27}\frac{d}{dx}\left[[3x-4]^2[3x^2+8x+16]\frac{d\Phi}{dx}\right]-\ell(2n+\ell-2)x^2\Phi=0.
\eea
The horizon is located at $x=4/3$, where the coefficient in front of the second derivative vanishes as $(3x-4)^2$. 
Motivated by solutions in five and seven dimensions, we make a change of variables
\bea
\Phi(x)=(3x-4)^p z^q g[z],\quad z=\frac{3x-4}{\nu_1 x+\nu_2}
\eea
and require function $g[z]$ to satisfy the standard hypergeometric equation with singular points at $z=0,1$. Substituting function 
$\Phi$ in equation (\ref{Extrm9dApp}), we find that the result has a schematic form
\bea
z^2(3-\nu_1 z)^2 Q_2 g''+z(3-\nu_1 z) Q_3 g'+ Q_4 g=0.
\eea
Here $Q_s$ are polynomials of degree $s$ in variable $z$. The last relation may reduce to the hypergeometric equation only if the following conditions are satisfied:
\bea
Q_4[0]=0&&\Rightarrow\quad 24(p_1+p_2)(p_1+p_2+1)=\ell(2n+\ell-2),\nn
Q_4[\frac{3}{\nu_1}]=0&&\Rightarrow\quad (4\nu_1+3\nu_2)^2[4p_1(3+p_1)-\ell(2n+\ell-2)]=0,\\
Q_3[\frac{3}{\nu_1}]=0&&\Rightarrow\quad (4\nu_1+3\nu_2)^2(1+p_1)=0.\nonumber
\eea
Since the option $(4\nu_1+3\nu_2)=0$ removes $z$ as an independent variable, the last two relations imply that
\bea
p_1=-1,\quad \ell(2n+\ell-2)=-8.
\eea
These are necessary conditions, which also turn out to be sufficient. Specifically, they imply that equation (\ref{Extrm9dApp}) becomes (\ref{PhiXeqnAextr9d}), and its solution (\ref{9dsolnHyper}) is indeed hypergeometric.

\subsection{General excitations of extremal black holes}
\label{AppExtrBC}

Let us look at equation (\ref{CyclEqnTwo}) in the extremal case. The mass and horizon radius are given by
\bea
M=n\left[\frac{na^2}{n-1}\right]^{n-1},\quad r_h=\frac{a}{\sqrt{n-1}}
\eea
Then the equation reads
\bea\label{CyclEqnTwoExtr}
r\frac{d}{dr}\left[\frac{\Delta}{r}\frac{d\Phi}{dr}\right]+\frac{(r^2+a^2)^{2n-2}-a^{2n-4}\Delta}{\Delta}
(ap)^2\Phi-\Lambda r^2 (r^2+a^2)^{n-2}\Phi=0.
\eea
To analyze the behavior of this equation in the vicinity of the horizon, we introduce a new variable 
\bea
z=r-r_h\,.
\eea
Then  in the vicinity of the horizon the equation becomes
\bea
&&\frac{9 a^2 n}{6 (n_-)^2} \Big[a^2  p^2 n+ 5 a \sqrt{n_-}  p^2 n z - 
    2 (n_-) (c_2 + 2 (1 - 3 n) n_1^2) z^2\Big] \Phi\nn
&&\quad+ 
 z^3 \Big[12 a^2 n^2 + 
8 a \sqrt{n_-} n (1 + 7 n)z + (8 - 20 n - 125 n^2 + 
       137 n^3) z^2\Big] \Phi'\\
&&\quad+ \frac{z^4}{3} \Big[18 a^2 n^2 + 
    6 a \sqrt{n_-}  n (4 + 13 n)z + (16 - 104 n - 79 n^2 + 167 n^3) z^2\Big] \Phi''=0\nonumber
\eea
Here we defined $n_-=n-1$. In each square bracket we dropped the terms of the order $\OO(z^3)$, and the derivatives are taken with respect to $z$. To solve this equation order--by--order in $z$, it is convenient to temporarily rescale the independent variable and to redefine the function as
\bea
z=\eps {\bar z},\quad \Phi(z)={\bar\Phi}({\bar z}).
\eea
Rearranging the terms according to powers of $\eps$, we find
\bea
0&=&
9 a^4 n^2  p^2 {\bar\Phi} + 
  36\eps^2 (a n n_-)^2 {\bar z}^3 \left[2 {\bar\Phi}' +  {\bar z}{\bar\Phi}''\right]\nn
&&+\eps \sqrt{n_-}\Big\{45 a^3 (n n_1)^2 {\bar z} {\bar\Phi} + 
 12\eps^2 n(n_-)^2 a{\bar z}^4\left[4  (1 + 7 n)   {\bar\Phi}'
+  (4 + 13 n) {\bar z} {\bar\Phi}''\right]\Big\}\nn
&&-18 \eps^2 a^2 n n_- (c_2 + 2 (1 - 3 n) p^2)  {\bar z}^2 {\bar\Phi} \\
&&\quad+2 \eps^4 (n_-)^3  {\bar z}^5\left[
 3 (-8 +12 n + 137 n^2){\bar\Phi}' + 2  
    (-16 +88 n  + 167 n^2)  {\bar z} {\bar\Phi}''\right]\nonumber
\eea
In the leading order in $\eps$ the equation truncates to
\bea
9 a^4 n^2  p^2 {\bar\Phi} + 
  36\eps^2 (a n n_-)^2 {\bar z}^3 \left[2 {\bar\Phi}' +  {\bar z}{\bar\Phi}''\right]\simeq 0,
\eea
and its solution reads
\bea\label{PhiNdepOne}
{\bar\Phi}\simeq c_1 \exp\left[\frac{ia p}{2 \eps (n-1) {\bar z}}\right] +c_2
\exp\left[-\frac{ia p}{2 \eps (n-1) {\bar z}}\right]
\eea
As we will explain below, only one of these terms should be kept to enforce normalizability at the horizon. This is well know for a more general case of a nontrivial time dependence, where one keeps only the outgoing modes $\exp[-i\omega t+i h(r)]$ with positive $h$. We are looking at the $\omega=0$ solutions, so either $\exp[i h(r)]$ or $\exp[-i h(r)]$ can be kept, but not both. We will focus on $c_2=0$. 

Expanding function ${\bar\Phi}$ in powers of $\eps$ as\footnote{Rather than putting  functions $({\bar\Phi}_1,{\bar\Phi}_2,\dots)$ in the exponent we wrote them in a separate factor. This parameterization simplies the expresssion for  $({\bar\Phi}_1$ and 
${\bar\Phi}_2$ written bellow.}
\bea\label{NormAnlzPhi}
{\bar\Phi}=\exp\left[\frac{1}{\eps}{\bar\Phi}_{-1}+ {\bar\Phi}_{0}\right]\left(1 +
   \eps {\bar\Phi}_1+\eps^2 {\bar\Phi}_2+\dots\right),
\eea
we arrive a sequence of equations for ${\bar\Phi}_k$. Relation (\ref{PhiNdepOne}) determines ${\bar\Phi}_{-1}$, and the next order gives ${\bar\Phi}_{0}$:
\bea\label{PhiNdepTwo}
{\bar\Phi}_{-1}=\frac{ia p}{2 (n-1) {\bar z}},\quad {\bar\Phi}_0=-\frac{(n-1) p}{6 \sqrt{n_-} n}\log{\bar z}\,.
\eea
At the next two orders in $\eps$, we find 
\bea\label{PhiNdepTwo1}
{\bar\Phi}_{1}=\frac{[36 i \lambda n +36n(2-n)\sqrt{n_-}p + i(5n^2 +  64 n - 40) p^2] {\bar z}}{72 a n^2 p}
\eea
and
\bea\label{PhiNdepTwo2}
{\bar\Phi}_2&=&-\frac{{\bar z}^2}{10368 a^2 n^4 p^2}\Big[1296 \lambda^2 n^2
+64 p n^3 (-162 i \sqrt{n_-} + 243 p + 
          621 i \sqrt{n_-} p^2 + 10 p^3) 
\Big.\nn
&& - 
    72 \la n [72 n^3 + 16 n (3 i \sqrt{n_-} - 4 n_1) p + 
       40 n_1^2+ n^2 (-72 - 204 i \sqrt{n_-} p  - 5 p^2)]\nn
&&+512 i n (3 \sqrt{n_-} + 10 i p) p^3
   + 1600 p^4 +    1296 n^5 (-4 i \sqrt{n_-} + p)p \\
&&+ 
       48 n^2 p^2 (-108 - 312 i \sqrt{n_-} n_1 + 77 p^2) \nn
   &&\Big.    +n^4 (15552 i \sqrt{n_-} - 11664 p - 
          6600 i \sqrt{n_-} p^2 + 25 p^3)p\Big]\nonumber
\eea
To identify the normalizable solutions, we look at an effective action corresponding to equation (\ref{CyclEqnTwo}),
\bea
S=\int dr \left[\frac{\Delta}{r}\left|\frac{d\Phi}{dr}\right|^2-\left\{\frac{(r^2+a^2)^{2n-2}-a^{2n-4}\Delta}{r\Delta}
(ap)^2\Phi-\Lambda r (r^2+a^2)^{n-2}\right\}|\Phi|^2\right]
\eea
Expansion in the vicinity of the horizon gives
\bea\label{NormAction}
S&=&\int dz\left\{ 4 (n-1)^2(1 + a_1 z + a_2 z^2) z^2 \left|\frac{d\Phi}{dz}\right|^2  - 
 a^2 p^2 (1 + a_3 z + a_4 z^2) \frac{|\Phi|^2}{z^2}\right\}\nn
&&+\Lambda \int dz \frac{2(n-1)}{n}|\Phi|^2
\eea
The values of constants $a_k$ are not very illuminating, but we give them for completeness:
\bea
&&a_1 = \frac{2 (n-2) \sqrt{n_-}}{3 a n}, \
 a_2 = \frac{(n-1)(24 -28 n + 11 n^2)}{12 a^2 n^2},\\
&&a_3 = \frac{4(n-2) \sqrt{n_-}}{3 a n}\,,\quad
 a_4=\frac{n_- (232-292 n+43 n^2)}{36 a^2 n^2}\nonumber
\eea
Introducing variable ${\bar z}=\eps^{-1} z$ and using the dominant contribution (\ref{PhiNdepOne}) to the wavefunction, we can write the leading contribution to (\ref{NormAction}) as
\bea\label{NormActionOne}
S&=&\frac{1}{\eps}\int d{\bar z}\left\{ 4(n-1)^2{\bar z}^2 \left|\eps^2\frac{d{\bar\Phi}}{d{\bar z}}\right|^2- 
 a^2 p^2  \frac{|{\bar\Phi}|^2}{{\bar z}^2}\right\}+\OO(\eps^0)\nn
&=&\frac{1}{\eps}\int d{\bar z}\left\{ \frac{a^2 p^2}{{\bar z}^2} \left|c_1 \zeta-c_2\zeta^{-1}\right|^2- 
   \frac{a^2 p^2}{{\bar z}^2}\left|c_1 \zeta+c_2\zeta^{-1}\right|^2\right\}+\OO(\eps^0)
\eea
Here
\bea
\zeta= \exp\left[\frac{ia p}{2 \eps (n-1) {\bar z}}\right].\nonumber
\eea
The divergent terms in (\ref{NormActionOne}) cancel if and only if the product $c_1 c_2$ vanishes, so the normalizable solution is given by (\ref{NormAnlzPhi})--(\ref{PhiNdepTwo}) which corresponds to $c_2$ equal to zero. By setting $c_1=0$ instead one gets a complex conjugate of that solution, but this is a discrete choice rather than a freedom in a continuous parameter. Substituting solution (\ref{NormAnlzPhi})--(\ref{PhiNdepTwo2}) into the action (\ref{NormActionOne}) and expanding the result in powers of $\eps$, one finds the leading non--vanishing contribution:
\bea
S=\eps^2  \int {\bar z}d{\bar z}(n_-)^{3/2}\left[\frac{2 {\Lambda} (11 n+8) }{3 a n^2}
-\frac{P p^2}{27 \left(a n^3\right)}-\frac{4n_-(n-2) }{a n}\right]
+O(\eps^3)\nn
\eea
Here $P= (2 n-1) \left(67 n^2-256 n-296\right)$. Convergence of this integral demonstrates normalizability of the solution (\ref{NormAnlzPhi})--(\ref{PhiNdepTwo2}).

\section{Technical details for the WKB approximation}
\label{AppWBK}

In section \ref{SecWKB} we found the expression (\ref{WkbLove}) for the Love numbers in the WKB approximation. Derivation of this result relied on matching wavefunctions between the 
near--horizon and the WKB regimes, and in the overlapping region we used the regularized asymptotics (\ref{WkbBeslAsmp}) of the Bessel functions. The resulting expression (\ref{WkbLove}) contained the parameter $\sigma$, which depended on the details of regularization but not on the quantum number $\ell$, and this property was sufficient for extracting the asymptotic behavior of the Love numbers. An unambiguous determination of $\sigma$ would involve introduction of a third region overlapping with both the near--horizon and the WKB regimes, and such analysis goes beyond the scope of our paper. Instead of performing such calculation, in this appendix we will demonstrate how the WKB approximation is recovered from the exact solutions, when those are known, and extract some insights into the nature of parameter $\sigma$. We will focus on two examples: static solutions in all odd dimensions and rotating geometries in $d=5$.

\bigskip
\noindent
{\bf Love numbers for static geometries}

We begin with analyzing exact solutions of equation  (\ref{ScalEqnSS}) describing regular solutions of a scalar field in static geometries and comparing the resulting wavefunctions (\ref{ScalEqnSS}) with their WKB approximation (\ref{eqn49})--(\ref{eqn410}). 

As demonstrated in section \ref{SecStatic}, the solution of equation (\ref{ScalEqnSS}) regular at the horizon is given by (\ref{PhiHyper})
\bea
\Phi=c_1 r^{\ell}F\left[-\frac{\ell}{d-3},-\frac{\ell}{d-3},1,1-\frac{M}{r^{d-3}}\right]\,,
\eea
At infinity, this function has three qualitatively different types of behavior depending on the value of $k=\frac{\ell}{d-3}$, and making comparison to the WKB approximation we will focus on the generic case where $k$ is neither integer nor half integer. Then the wavefunction can be rewritten as (\ref{HyperAC}):
\bea\label{PhiHyperWKB}
\Phi=c_1 r^{\ell}\left[\frac{\Gamma[1+2k]}{\Gamma[1+k]^2}F[-k,-k,-2k,z]+
\frac{\Gamma[1-2p]}{\Gamma[1-p]^2}z^{2k+1}F[p,p,2p,z]\right],
\eea
where $z=\frac{M}{r^{d-3}}$ and $p=k+1$. The WKB approximation becomes applicable if 
$k\gg 1$, and in this case the hypergeometric function can be approximated by
\bea
F[p,p,2p,z]=\frac{1}{(1-z)^{\frac{1}{4}}}
\left[\frac{1}{2}+\frac{\sqrt{1-z}}{2}\right]^{1-2p}\left[1+\frac{f_1(z)}{p}+
\frac{f_2(z)}{p^2}\dots\right]
\eea
and a similar expression with the replacement $p\rightarrow (-k)$. Substituting the leading contributions into (\ref{PhiHyperWKB}) and using the Stirling approximation for gamma functions, we find
\bea\label{PhiHyperWKBapp}
\Phi\simeq \frac{{\tilde c}_1 r^{\ell}}{(1-z)^{\frac{1}{4}}}
\left[W^{2k+1}+
2^{-4k-3}\tan(\pi k) z^{2k+1}W^{-2k-1}\right],\quad
W=\frac{1}{2}+\frac{\sqrt{1-z}}{2}
\eea
Here we rescaled the overall coefficient $c_1$, and used the relation
\bea
\frac{\Gamma[1-2p]}{\Gamma[1-p]^2}\frac{\Gamma[1+2k]}{\Gamma[1+k]^2}=
\frac{(k+\frac{1}{2})\tan(\pi k)}{\pi}\frac{\Gamma[k+1]^4}{\Gamma[2(k+1)]^2}
\simeq 2^{-4(k+1)+1}\tan(\pi k)
\nonumber
\eea
Let us now compare (\ref{PhiHyperWKBapp}) with the WKB result (\ref{eqn49})--(\ref{eqn410}).
Setting $a=0$ in (\ref{eqn410}) and evaluating the integral, we find
\bea
S_0=\frac{1}{2}\ln\frac{x}{x_h}+\frac{1}{n-1}\ln\left[1+\sqrt{1-z}\right],\quad 
z=\left[\frac{x}{x_h}\right]^{n-1}
\eea
This leads to the asymptotic expression for (\ref{eqn49})-(\ref{eqn410}) at large values of $x$:
\bea
\Phi&\simeq& {\bar c}_1 \left[\frac{1}{1-M x^{1-n}}\right]^{\frac{1}{4}}x^{\ell/2}
\left\{(2W)^{2k+1}+\sigma \left[\frac{x}{x_h}\right]^{-(\ell+n-1)}
(2W)^{-2k-1}\right\}\nn
&\simeq& {\tilde c} \left[\frac{1}{1-z}\right]^{\frac{1}{4}}2^{-2k-1}r^\ell 
\left\{W^{2k+1}+\sigma\, 2^{-4k-2} z^{2k+1}
W^{-2k-1}\right\}
\eea
This expression is in the perfect agreement with (\ref{PhiHyperWKBapp}), and as expected, parameter $\sigma$ is of order one. Both expressions lead to the Love numbers given by (\ref{WBBstat1}).

\bigskip
\noindent
{\bf Love numbers for rotating black holes in five dimensions}

Let us now consider the exact solution of equation for scalar field in five dimensional black holes with equal rotation parameters. As discussed in Appendix \ref{App5d}, the solution is given by 
\bea
\Phi=c_1 z^{-\ell/2}F\left[-\frac{\ell}{2},-\frac{\ell}{2},1,1-z\right]\,,
\eea
where
\bea
z^{-1}=\frac{1}{\sqrt{M(M-4a^2)}}\left[r^2-\frac{M}{2}+a^2\right]+\frac{1}{2}\,.
\eea
Repeating the arguments which led to (\ref{PhiHyperWKBapp}), we arrive at the same expression with the replacement
\bea\label{WKBrRepl}
r^\ell\rightarrow z^{-\ell/2}\,.
\eea
Let us compare this result with the WKB approximation (\ref{PhiWKBfin})--(\ref{eqn410}) for $n=2$. The integral in (\ref{eqn410}) gives
\bea
S_0(x)=\int_{x_h}^x \frac{\sqrt{{x_h-a^2}}x dx}{2x\left[(x_h-a^2)x^2-(x-a^2)(x_h)^2\right]^{\frac{1}{2}}}=
-\int_1^z \frac{dz}{2z\sqrt{1-z}}=\ln\left[\frac{W}{\sqrt{z}}\right]
\eea
with $W$ defined in (\ref{PhiHyperWKBapp}). We used the relation
\bea
x^2-\frac{(x_h)^2}{(x_h-a^2)}(x-a^2)=(r^2+a^2)^2-M r^2=M(M-4a^2)z^{-1}(z^{-1}-1)
\nonumber
\eea
and recalled that
\bea
x_h=r_h^2+a^2=\frac{1}{2}\left[M+\sqrt{M-4a^2}\right].\nonumber
\eea
The prefactor in equation (\ref{eqn49}) reads
\bea
\frac{1}{x^2-M x+M a^2}=\frac{1}{M(M-4a^2)}\frac{z^2}{1-z}\,,
\eea
and it leads to the final expression for the wavefunction in the WKB approximation:
\bea
\Phi&\simeq&  {\tilde c} \left[\frac{1}{1-z}\right]^{\frac{1}{4}}2^{-2k-1}z^{\ell/2}
\left\{W^{2k+1}+\sigma\, 2^{-4k-2} z^{2k+1}
W^{-2k-1}\right\}\,.
\eea
Since neither this relation nor (\ref{PhiHyperWKBapp}) with replacement (\ref{WKBrRepl}) contain the rotation parameter, the perfect agreement between them follows from the analysis of the static case carried out above.


\end{document}